%% file: Bethe-SalpeterCJP.tex
\documentclass{cjp}
\usepackage{amsmath,amsbsy,amsfonts}
\newcommand{\nn}{\nonumber \\}
\newcommand{\eps}{\ensuremath{\varepsilon}}
\newcommand{\me}{\mathrm{e}}
\newcommand{\im}{\ensuremath{\mathrm{i}}}
\newcommand{\bra}[1]{\langle #1 |}
\newcommand{\ket}[1]{| #1 \rangle}

\newcommand{\bigbra}[1] {\big\langle #1\big|}
\newcommand{\bigket}[1] {\big|#1\big\rangle}
\newcommand{\Bigbra}[1] {\Big\langle #1\Big|}
\newcommand{\Bigket}[1] {\Big|#1\Big\rangle}
\newcommand{\dif}{\ensuremath{\mathrm{d}}}

\newcommand{\bx}{\boldsymbol{x}}

\newcommand{\Partder}[1]{\frac{\partial }{\partial #1}}

\newcommand{\partder}[2]{\frac{\partial #1}{\partial #2}}
\newcommand{\deriv}[2]{\frac{\dif #1}{\dif #2}}
\newcommand{\partdelta}[2]{\frac{\delta #1}{\delta #2}}
\newcommand{\ppartder}[2]{\frac{\partial^2 #1}{\partial^2 #2}}
\newcommand{\ppartdelta}[2]{\frac{\delta^2 #1}{\delta^2 #2}}
\newcommand{\pppartdelta}[2]{\frac{\delta^3 #1}{\delta^3 #2}}
\newcommand{\pppartder}[2]{\frac{\partial^3 #1}{\partial^3 #2}}

\newcommand{\npartder}[3]{\frac{1}{#1!}\,{\frac{\partial^#1 #2}
{\partial #3^#1}}}

\newcommand{\partdern}[3]{{\frac{\partial^#1 #2}
{\partial #3^#1}}}
\newcommand{\Cov}{\mathrm{Cov}}
\newcommand{\eff}{_{\mathrm{eff}}}

\newcommand{\sgn}{\mathrm{sgn}}

\renewcommand{\rm}{\mathrm}
\newcommand{\mI}{\mathrm{I}}

\newcommand{\mD}{\mathrm{D}}

\newcommand{\mH}{\mathrm{H}}

\newcommand{\conn}{\mathrm{conn}}

\newcommand{\bs}{\boldsymbol}

\renewcommand{\H}{H}

\newcommand{\dagg}{^{\dag}}
\newcommand{\intd}[1]{\int\frac{\dif #1}{2\pi}}
\newcommand{\half}{{\displaystyle\frac{1}{2}}}
\newcommand{\halfS}{{\scriptstyle\frac{1}{2}}}
\newcommand{\tref}{{\displaystyle\frac{1}{3!}}}

\newcommand{\dint}{\int\!\!\!\int}
\newcommand{\ddint}{\dint\!\!\!\dint}

\newcommand{\intbx}{\int\dif^3\bx}
\newcommand{\vsp}{\vspace{0.5cm}}

\newcommand{\Vsp}{\vspace{1cm}}

\newcommand{\hsp}{\hspace{0.5cm}}
\newcommand{\Hsp}{\hspace{1cm}}

\newcommand{\hpsi}{\hat\psi}

\renewcommand{\and}{\quad\mathrm{and}\quad}
\renewcommand{\it}{\textit}

\newcommand{\abs}[1]{|{#1}|}

\newcommand{\eq}{\eqref}
\newcommand{\rarr}{\rightarrow}

\newcommand{\Rarr}{\Rightarrow}

\newcommand{\rr}{r_{12}}

\newcommand{\DF}{D_{\rm{F}\mu\nu}}
\newcommand{\SF}{S_{\rm{F}}}

\newcommand{\calH}{\mathcal{H}}

\newcommand{\calKb}{\kappa}

\newcommand{\calV}{\mathcal{V}}
\newcommand{\V}{\mathcal{V}}
\newcommand{\calE}{\mathcal{E}}
\newcommand{\Q}{\mathcal{\boldsymbol{Q}}}
\newcommand{\gamlim}{\ensuremath{\gamma\rightarrow 0}}
\newcommand{\limgam}{\lim_{\gamlim}}
\newcommand{\Ugam}[1]{U_\gamma(#1,-\infty)}
\newcommand{\Ugamtil}[1]{\widetilde{U}_\gamma(#1,-\infty)}
\newcommand{\Ugamt}{\widetilde{U}_\gamma}
\newcommand{\Util}{\widetilde{U}}

\newcommand{\TD}{T_\rm{D}}

\newcommand{\psiH}{\hat{\psi}_\mH}
\newcommand{\psiI}{\hat{\psi}_\mI}
\newcommand{\epsi}{\epsilon}
\newcommand{\Ombar}{\bar{\Omega}}
\newcommand{\Om}{\Omega}
                             \newcommand{\bsdot}{\bs{\cdot}}

                                  \newcommand{\G}{\Gamma}
                                  \newcommand{\GQ}{\Gamma_Q}
                                   \newcommand{\GP}{\Gamma_P}
                             
\begin{document}
\title{Many-body-QED perturbation theory: Connection to the Bethe-Salpeter equation}
\author{Ingvar Lindgren}
\address{Department of Physics, Chalmers University of Technology and
the G\"oteborg University, G\"oteborg, Sweden}
\shortauthor{Lindgren} \maketitle

\begin{abstract} The connection between many-body theory (MBPT)---in perturbative
and non-perturbative form---and quantum-electrodynamics (QED) is
reviewed for systems of two fermions in an external field. The
treatment is mainly based upon the recently developed
covariant-evolution-operator method for QED calculations [Lindgren
\it{et al.} Phys. Rep. \textbf{389}, 161 (2004)], which has a
structure quite akin to that of many-body perturbation theory. At
the same time this procedure is closely connected to the
$S$-matrix and the Green's-function formalisms and can therefore
serve as a bridge between various approaches. It is demonstrated
that the MBPT-QED scheme, when carried to all orders, leads to a
Schr\ödinger-like equation, equivalent to the Bethe-Salpeter (BS)
equation. A Bloch equation in commutator form that can be used for
an "extended" or quasi-degenerate model space is derived. It has
the same relation to the BS equation as has the standard Bloch
equation to the ordinary Schr\ödinger equation and can be used to
generate a perturbation expansion compatible with the BS equation
also for a quasi-degenerate model space.
\\\\PACS Nos.: {31.10+z, 31.15Md, 31.30Jv}
\end{abstract}
\begin{resume}French version of abstract
(supplied by CJP) \traduit\end{resume}

\begin{center}Submitted 25 Jan. 2005, Corrected 8 Feb. 2005\end{center}

\input{FigurecommandsV.tex}
\setlength{\unitlength}{0.75cm}

\section{Introduction}
\subsection{General} What is known as the Bethe-Salpeter (BS) equation
represents the complete solution of the relativistic two-body
problem with important applications in various branches of
physics. The equation was first derived by Bethe and Salpeter in
1951~\cite{SB51}, using the relativistic $S$-matrix formalism and
the analogy with Feynman graphs, and at about the same time by
Gell-Mann and Low~\cite{GML51}, using a rigorous field-theoretical
approach based on Green's functions. A closely related equation
was discussed by Schwinger in his Harvard lectures already in the
late 1940's~\cite{Schw51,KarpK52,Schw64,Nam97}.

In interpreting the solutions of the BS equation, several serious
problems were encountered, as discussed early by
Dyson~\cite{Dyson53}, Wick~\cite{Wick53} and
Goldstein~\cite{Gold53}. Dyson was particularly concerned about
the meaning of the wave function in relativistic quantum
mechanics, a subject \it{"full of obscurities and unsolved
problems"}. Solving the BS equation leads to a 4-dimensional wave
function---with individual times for the two particles. This
function is manifestly relativistically covariant but not in
accordance with the standard quantum-mechanical picture. That
leads to "spurious" or "abnormal" solutions without physical
significance and with no nonrelativistic
counterpart~\cite{Sazd87}. Another fundamental problem is that the
BS equation does not reduce to the correct "one-body limit", when
one of the particles becomes infinitely heavy, as discussed by
Gross and others~\cite{Gross69,Gross82}. Problems of these kinds
are most pronounced in the scattering of strongly interacting
particles but less so for bound-state systems in
weak-coupling~\cite{Tod71,CasLep78,Conn91,PhilWal96,Bijt01} (see
ref. \cite{Nam97} for a review).

The earliest applications of the BS equation appeared in atomic
physics and concerned the proton recoil contribution to the
hydrogen fine structure by Salpeter~\cite{Salp52} and the
positronium energy level structure by Karplus and
Klein~\cite{KarpK52}.

An important goal for the equation has been the study of
\it{strongly interacting particles}, which is a fundamental
problem in elementary-particle physics. In recent years there have
been numerous applications in QCD, dealing mainly with the
quark-quark, quark-antiquark interactions, quark confinement and
related problems~\cite{CasLep78,LSG91,BRS96,MR97}. Here, the
problems mentioned above are more serious, as recently summarized
by Namyslowski~\cite{Nam97}.

There have also been many applications in surface and solid-state
physics, ranging from electron-hole interactions in ion
crystals~\cite{ORR02} and studies of the two-dimensional Hubbard
model~\cite{BSW89} and Cooper pairs~\cite{WGH02} to quantum
dots~\cite{AFL04}.

 The BS equation has also been applied to three
or more particles~\cite{KL03,WR00,Taylor66}, although serious
problems have been encountered for more than three
particles~\cite{Bijt02}.

Various approximation schemes for treating the BS equation have
been developed over the time. The simplest approximation is the
\it{"ladder approximation"}, where all intermediate states evolve
only in the forward (positive) time direction. This is a useful
starting point in the strong-coupling case, where the standard
perturbative or self-consistent approach may not converge, and
this approximation is, for instance, the basis for the Brueckner
theory of nuclear matter~\cite[Sect. 41]{Brueck59,FW71}. Another
approach is the \it{"quasi-potential approximation"}, which
implies that the equation is reduced to an equivalent
3-dimensional Schr\ödinger equation, which can be done without
loosing any rigor~\cite{CasLep78,PTjon98}. Early numerical
calculations in the this regime were done particularly by Schwartz
and Zemach~\cite{SchwZ66} and Kaufmann~\cite{Kaufm69}.

In atomic physics the BS equation has been applied mainly in
treating positronium~\cite{AFM02,AF99} and to heliumlike ions, and
we shall be particular concerned with the latter here. This is
strictly speaking a three-body problem but can to a good
approximation be treated---with the first Born approximation---as
a two-body problem with an external potential. The application to
heliumlike systems was pioneered by Sucher~\cite{Su57a,Su58} and
Araki~\cite{Ar57} in the late 1950's for deriving the leading
relativistic and QED energy corrections beyond the Breit
interaction. Later these works have been extended---largely along
the lines of Sucher---by Douglas and Kroll in the
1970's~\cite{DK74} and more recently by Zhang and
Drake~\cite{ZDr94,ZDr96,Zhang96,Zhang96a}.

The technique developed by Drake and coworkers is presently the
most accurate available in dealing with heliumlike systems. The
wave functions used are very accurate functions of Hylleraas type,
and the QED corrections are evaluated by means of analytical
expressions up to order $\alpha^5$ Ry (atomic units, or
$m\alpha^7$ in relativistic units), derived from the BS equation.
The wave functions used by Drake et al. are nonrelativistic but
certain relativistic effects are treated to all order in the
"\it{unified model}"~\cite{Dr79,Dr88}. The analysis of the BS
equation are in these works based upon the Brillouin-Wigner
perturbation theory (BWPT).

\newcommand{\Rect}[3]
{\put(0,0){\LineH{#1}} \put(0,#2){\LineH{#1}}
\put(0,0){\makebox(#1,#2){ #3}} \put(0,0){\LineV{#2}}
\put(#1,0){\LineV{#2}}}

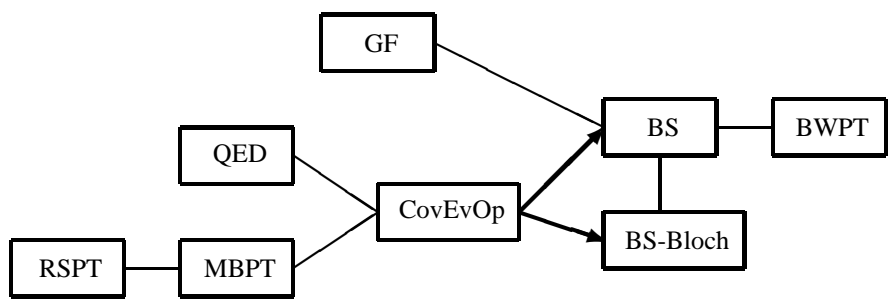
\begin{figure}[htb]
\begin{center}
\thicklines
\begin{picture}(5,5)(0,1.5)
  \put(-4,1.5){\line(1,0){1}}
  \put(-1,3.5){\line(3,-2){1.5}}
  \put(-1,1.5){\line(3,2){1.5}}
  \put(3,2.46){\line(1,1){1.35}}
  \put(4.48,4){\vector(1,1){0}}
  \put(4.52,3.96){\vector(1,1){0}}
  \put(3,2.5){\line(1,1){1.5}}
  \put(3,2.52){\line(1,1){1.3}}
  \put(3,2.48){\line(1,1){1.3}}
  \put(3,2.5){\line(3,-1){1.5}}
  \put(3,2.52){\line(3,-1){1.5}}
  \put(3,2.48){\line(3,-1){1.5}}
  \put(4.49,1.96){\vector(3,-1){0}}
  \put(4.52,2.02){\vector(3,-1){0}}
  \put(6.5,4){\line(1,0){1}}
  \put(1.5,5.5){\line(2,-1){3}}
  \put(5.5,2.5){\line(0,1){1}}
  \put(-0.5,5){\Rect{2}{1}{GF}}
  \put(-6,1){\Rect{2}{1}{RSPT}}
  \put(-3,1){\Rect{2}{1}{MBPT}}
  \put(-3,3){\Rect{2}{1}{QED}}
  \put(0.5,2){\Rect{2.5}{1}{CovEvOp\large }}
  \put(4.5,3.5){\Rect{2}{1}{BS}}
  \put(4.5,1.5){\Rect{2.5}{1}{BS-Bloch}}
  \put(7.5,3.5){\Rect{2}{1}{BWPT}}
 \end{picture}
   \Vsp\caption{Schematic illustration of the connection between various many-body techniques.
   The upper right part represents the Green's-function (GF) approach, which is used to
   derive the Bethe-Salpeter (BS) equation, normally analyzed in terms of the Brillouin
   Wigner perturbation theory (BWPT). The lower-left part illustrates the many-body
   perturbation theory (MBPT), originating from Rayleigh-Schr\ödinger perturbation
   theory (RSPT). The combination of quantum-electrodynamics (QED)
   with MBPT is represented by the covariant-evolution-operator
   (CovEvOp) method, and the link to the BS equation and the corresponding Bloch equation---the main
   subject of the present paper---is illustrated by the arrows.}
   \label{Fig:Intr}
 \end{center}
 \end{figure}

A different and in some aspects more versatile approach to the
many-body problem is the procedure known as the \it{many-body
perturbation theory} (MBPT). This is based upon
Rayleigh-Schr\ödinger perturbation theory (RSPT)~\cite{LM86},
which via the Bloch equation can be used to derive various
computational schemes, such as the \it{linked-diagram expansion}
(LDE)~\cite{Br55,Go57,Br67}. A particularly powerful technique is
the \it{Coupled-Cluster Approach} (CCA)~\cite{CK60,Ci66,PC75},
which is widely used in quantum chemistry~\cite{BP78,KB92CC}. This
technique is non-perturbative but closely connected to MBPT, and
we shall include it in the MBPT category here. The MBPT techniques
are primarily developed for the weak-coupling case, but might in
the non-perturbative (CCA) form be used also in strong coupling.

The MBPT procedures, based initially upon RSPT, have the great
advantage compared to techniques based upon BWPT that they are
\it{size-extensive} at each order, which implies that the energy
scales linearly with the size of the system---a property of vital
importance for molecular problems~\cite{PKSB78,BSP79}. Both
procedures can also be combined with the \it{extended-model-space
technique}, which is particularly effective in dealing with
problems of \it{quasi-degeneracy}~\cite{Li74,Li78,LM86,MLL95}.

For QED problems the \it{$S$-matrix technique} has been the
standard procedure since the days of Feynman and Dyson. (For a
review of the application to bound-state problems, see
ref.~\cite{MPS98}.) Being based upon scattering theory, this
technique has the disadvantage that its structure is quite
different from that of MBPT, which makes it hard to combine the
procedures (see, e.g. ref.~\cite{SB89}). The standard procedure
for such a combination has been to perform a separate
(relativistic) many-body calculation and adding first-order energy
corrections from QED analytically~\cite{PJS94}. This procedure
gives in many cases satisfactory results but is hard to improve in
any systematic way. In particular, it gives no additional
information about the wave function.

Another disadvantage with the $S$-matrix formalism is that the
energy is conserved between the initial and the final states. This
implies that it cannot be combined with the extended-model-space
technique, successfully applied in MBPT. This techniques requires
generally elements of the effective interaction that are
nondiagonal in energy. This problem has recently been remedied by
means of a new technique, known as the
\it{Covariant-Evolution-Operator method} (CovEvOp), which is a
modification of the standard evolution-operator technique of
time-dependent perturbation theory~\cite{FW71} in order to make it
applicable to relativistic problems (for a review, see
ref.~\cite{LSA04}). This technique has a structure that is very
akin to that of MBPT, and it deals with the key ingredients of
MBPT---the wave operator and the effective interaction. At the
same time the method is closely related to the $S$-matrix
formalism and the Green's-function procedure. The technique can
therefore be regarded as a merger of MBPT/CCA and QED~\cite{Li00},
and it has recently been successfully applied to the
quasi-degenerate fine-structure states of heliumlike
systems~\cite{LAS01}.

The quasi-degenerate problem can also be handled with the
\it{two-times Green's-function} approach, developed by Shabaev and
coworkers (for a review, see ref.~\cite{Shab02}). This technique,
however, has no direct link to MBPT and will therefore not be
discussed further here.

The procedure with the Covariant-Evolution-Operator method is now
being further developed at our laboratory in order to combine QED
and MBPT in a more complete fashion. This will be based on the
non-perturbative coupled-cluster approach (CCA) of electron
correlation or the so-called Dirac-Coulomb approximation,
corresponding to the "ladder approximation" of the Bethe-Salpeter
equation. This is combined with a perturbative expansion of the
remaining (mainly QED) effects, which in principle leads to the
full BS equation. This is along the lines early drawn by
Sucher~\cite{Su58} and followed by many later
works~\cite{DK74,Conn91,Zhang96,AF99,SauliA03}. Our approach
differ from all the earlier ones in the sense that all effects are
evaluated \it{numerically} rather than analytically.

Our approach implies that the QED effects are evaluated with
highly correlated (relativistic) wave functions, and for
two-electron systems the results will then, in principle, be
comparable to those of Drake's unified method, with the difference
that the relativistic effects are included in a complete way and
that the QED effects are evaluated numerically.

In the diagram in Fig. \ref{Fig:Intr} we have tried to represent
the relations between the many-body approaches described here in a
simple and illustrative way. The many-body procedures based upon
Rayleigh-Schr\ödinger perturbation theory are indicated in the
lower-left part and the Green's-function and Bethe-Salpeter
procedures, more associated to Brillouin-Wigner perturbation
theory, in the upper-right part. The present paper deals
particularly with the connection between the two approaches,
represented by the arrows in the diagram.

In addition to deeper insight into the different procedures, the
present treatment will make it possible to analyze a problem based
on the BS equation in terms of RS-MBPT---not only in terms of
BWPT, as has previously been the case~\cite{DK74,Zhang96a}. The
Bloch equation in commutator form, compatible with the BS
equation, which is derived, has the same relation to the BS
equation as has the standard Bloch equation to the ordinary
Schr\ödinger equation, and it could possibly be used to eliminate
the quasi-degeneracy problem that might appear when the BS
equation is treated for a single state at a time.

Since the equivalence of the MBPT-QED-CovEvOp procedure with the
BS equation has now been established for two-electron systems,
this new link will probably make it easier to apply the BS
procedure---or its equivalence--- also to systems with more
electrons. Alternatively, this can be used to analyze a
many-body-QED calculation to find out what is missing in order to
represent a complete Bethe-Salpeter treatment. Our main emphasize
here is applications to atoms and other weak-interacting systems.
Since the procedure we have developed, however, is based upon a
combination of perturbative and non-perturbative approaches, the
results obtained might be useful also outside this regime.

The paper will be organized in the following way. Below we shall
first conjecture the Bethe-Salpeter equation in a simple-minded
way as an introduction. In section \ref{sec:ConvPT} we shall
summarize the necessary ingredients of time-independent and
time-dependent perturbation theory and in the following section
briefly review the original derivations of the Bethe-Salpeter
equation by Bethe and Salpeter and by Gell-Mann and Low, based on
Green's functions. The main part of the paper will be devoted to a
rigorous derivation of the Bethe-Salpeter equation, starting from
the covariant-evolution-operator method. The basics of the method
are summarized in section \ref{sec:EvolOp}, and the method will
then be used to derive the Bethe-Salpeter equation. A
corresponding Bloch equation will also be derived, which will make
it possible to treat the BS equation perturbatively also for a
quasi-degenerate (extended) model space. Technical details of the
treatment are given in a number of appendices. Radiative effects
(self energies and vacuum polarization) are not considered here
but can be included by modifying the electron propagator and
photon interactions, as discussed, for instance, by Douglas and
Kroll~\cite{DK74}.

\subsection{Bethe-Salpeter equation}
An equation of BS type can be conjectured in a very simple way by
considering the time-independent nonrelativistic Schr\ödinger
equation
\begin{equation}
  \label{SchrEq}
  H\Psi=E\Psi
\end{equation}
with $H=H_0+V_1$, where $H_0=h_1+h_2$ is the zeroth-order
Hamiltonian (sum of single-electron Hamiltonians) and
$V_1=e^2/\rr$ is the electron-electron interaction (in
relativistic units \footnote{In this article relativistic units
are used, i.e., $m=c=\hbar=\epsi_0=1, e^2=4\pi\alpha$, where
$\alpha$ is the fine-structure constant.}). The Schr\ödinger
equation can then be expressed
\begin{equation}
  \label{Schr}
  (E-H_0)\Psi=V_1\Psi
\end{equation}
with the solution
\begin{equation}
  \label{SE1}
  \Psi=\Gamma(E)V_1\Psi,
\end{equation}
where
\begin{equation}
  \label{Gamma1}
  \Gamma(E)=\frac{1}{E-H_0}=\frac{\ket{rs}\bra{rs}}{E-\eps_r-\eps_s}
\end{equation}
is the \it{"resolvent"} operator~\cite[Ch. 9]{LM86} and $\ket{rs}$
is the Dirac notation of the straight (not antisymmetrized)
product of two single-electron functions, satisfying the Dirac
equation
\begin{equation}
  \label{SEeq}
  h\ket{i}=\eps_i\ket{i}
\end{equation}
We apply the summation convention, implying summation over
repeated indices appearing on one side of the equation. Unless
specifies otherwise, the summation is performed over positive-
(particle) as well as negative-energy (hole) states.

In the relativistic formalism one should, following
Sucher~\cite{Su57a,Su58}, replace $V_1$ by
$\Lambda_{++}e^2/\rr\Lambda_{++}$, where $\Lambda_{++}$ is the
projection operator for particle (positive-energy) states. This
leads to the \it{Coulomb-ladder approximation}, mentioned above,
i.e., a series of Coulomb interactions separated by particle
states. In QED $V_1$ can in the first approximation be replaced by
the \it{energy-dependent} interaction with a fully \it{covariant
photon} $V_1(E)$, i.e., Coulomb and transverse photon, the latter
representing (retarded) Breit interaction. In the next step
$V_1(E)$ can be replaced by $V_1(E)+V_2(E)$, where $V_2(E)$
represents the \it{non-separable} (irreducible) interaction of two
photons, i.e., the interaction of two covariant photons that in
the QED description cannot be represented by repeated
single-photon interactions (see Fig. \ref{Fig:NonSep} below).
Continuing this process, summing all non-separable interactions
with one, two, ... photons
\begin{equation}
  \label{calV}
  \calV(E)=V_1(E)+V_2(E)+\cdots
\end{equation}
leads to
\begin{equation}
  \label{BS1}
  \Psi=\Gamma(E)\,\calV(E)\Psi
\end{equation}
or
\begin{equation}
  \label{BS}
  (E-H_0)\,\Psi=\calV(E)\Psi
\end{equation}
This is equivalent to the Schr\ödinger-like form of the
\it{Bethe-Salpeter equation} derived by Sucher~\cite[Eq.
1.47]{Su58} and also used by Douglas and Kroll~\cite[3.26]{DK74}
and by Zhang~\cite[Eq. 15]{Zhang96}.

The BS equation \eq{BS} can be expanded in terms of a
Brillouin-Wigner perturbation series~\cite[Ch. 9]{LM86}
\begin{equation}
  \label{BSBW}
  \Psi=\Psi_0+\Big(\GQ(E)\,\calV(E)+\GQ(E)\,\calV(E)\GQ(E)\,\calV(E)
  +\cdots\Big) \Psi_0
\end{equation}
where $\Psi_0$ is the unperturbed wave function and
\begin{equation}
  \label{RedRes}
  \GQ(E)=\frac{Q}{E-H_0}
\end{equation}
is the "reduced" resolvent \eq{Gamma1} with the unperturbed state
removed. For this sequence to converge properly, it is required
that there be no eigenstate of $H_0$ close in energy to that of
$\Psi_0$ and of the same symmetry. A rigorous derivation of the
equation will be given in the following sections.

\section{Conventional many-body perturbation theory}
\label{sec:ConvPT}
\subsection{Time-independent perturbation theory}
\label{sec:MBPT} In time-independent many-body perturbation theory
(MBPT) (see, e.g., ref.~\cite{LM86}) the aim is to solve the
Schr\ödinger equation by successive approximations for a number of
\it{"target"} states
\begin{equation}
  \label{SE}
  H\,\Psi^\alpha(\bx)=E^\alpha\,\Psi^\alpha(\bx)\,;\qquad
  (\alpha=1,2,\cdots d)
\end{equation}
($\bx$ stands here for all space coordinates). The
time-independent Hamiltonian is partitioned into a zeroth-order
Hamiltonian and a perturbation
\begin{equation}
  \label{Part}
  H=H_0+H'
\end{equation}
For each target state $\Psi^\alpha(\bx)$ there exists a \it{model
state} or \it{zeroth-order wave function} (ZOWF)
$\Psi_0^\alpha(\bx)$ that is confined to a subspace, the \it{model
space} ($P$), spanned by eigenfunctions of $H_0$. The model space
can be degenerate or non-degenerate (quasi-degenerate). In the
latter case the model states are not necessarily eigenstates of
$H_0$. It is always assumed that all degenerate states of $H_0$
are either entirely inside or entirely outside the model space.

A \it{wave operator} $\Omega$ can be defined so that it transfers
all model states to the corresponding target states
\begin{equation}
  \label{Wop}
  \boxed{\Psi^\alpha(\bx)=\Omega\,\Psi_0^\alpha(\bx)\,; \qquad
  (\alpha=1,2,\cdots d)}
\end{equation}
In the following we shall use the \it{intermediate normalization}
(IN), implying that
\begin{equation}
  \label{IN}
  \big\langle\Psi_0^\alpha(\bx)\bigket{\Psi^\alpha(\bx)}=1
\end{equation}
The model states are the projections of the target states on the
model space
\begin{equation}
  \label{ZOWF}
  \Psi_0^\alpha(\bx)=P\Psi^\alpha(\bx)
\end{equation}
which implies
\begin{equation}
  \label{POP}
  P\Omega P=P
\end{equation}

The exact energies as well as the model states are obtained by
solving the secular equation
\begin{equation}
  \label{SecEq}
  H\eff\,\Psi_0^\alpha(\bx)=E^\alpha\,\Psi_0^\alpha(\bx),
\end{equation}
within the model space. Here, $H\eff$ is the \it{effective
Hamiltonian}, in IN given by
\begin{equation}
  \label{HeffIN}
  H\eff=PH\Omega P
\end{equation}

The wave operator satisfies the \it{generalized Bloch
equation}~\cite{Li74,LM86}
\begin{subequations}
\begin{equation}
  \label{Bloch}
  \boxed{\big[\Omega,H_0\big]P=\big(H'\Omega -\Omega\,
  H'\eff\big)P}
\end{equation}
where $H'\eff$ is the \it{effective interaction} (in IN)
\begin{equation}
  \label{EffInt2}
  H'\eff=H\eff-PH_0P=PH'\Omega P
\end{equation}
For a degenerate model space with the energy $E_0$ the equation
goes over into the original Bloch equation~\cite{Bl58a,Bl58b}
\begin{equation}
  \label{BlochDeg}
  (E_0-H_0)\,\Omega P=\big(H'\Omega -\Omega\,H'\eff\big)P
\end{equation}
\end{subequations}

The Bloch equation contains generally the information of a
\it{system} of Schr\ödinger equations \eq{SE}, corresponding to a
number of target states. The equation can conveniently be used as
the starting point for generating various perturbative and
non-perturbative schemes~\cite{Li74,LM86}. It leads directly to a
generalized form of the Rayleigh-Schr\ödinger perturbation
expansion, and it can be used to generate the \it{linked-diagram
expansion} (LDE) as well as the non-perturbative
\it{coupled-cluster approach} (CCA). The commutator form of the
Bloch equation \eq{Bloch} makes it possible to work with a
non-degenerate or "extended" model space", which is of particular
importance for quasi-degenerate problems, as mentioned above.

\subsection{Time-dependent perturbation theory}
\label{sec:TDPT} In time-dependent perturbation theory we start
from the time-dependent Schr\ödinger equation
\begin{equation}
  \label{TDSE}
  \im\Partder{t}\,\chi(t,\bx)=H(t)\,\chi(t,\bx)
\end{equation}
As before, $\bx$ stands for \it{all} space coordinates, while $t$
is a single time variable. Even if the Hamiltonian may be formally
time-dependent, we are interested in states that are
\it{stationary}, which implies that the wave function has the form
\begin{equation}
  \label{wft}
  \chi(t,\bx)=\Psi(\bx)\,e^{-\im Et}
\end{equation}
where $E$ is the energy of the system and $\Psi(\bx)$ is the
time-independent wave function. The latter is then a solution the
time-independent Schr\ödinger equation \eq{Schr}
\begin{equation}
  \label{Seq}
  H\,\Psi(\bx)=E\Psi(\bx)
\end{equation}

In the \it{interaction picture} (IP)~\cite{FW71} with the
partitioning \eq{Part} the wave function is related to that of the
Schr\ödinger picture by
\begin{equation}
  \label{IP}
  \chi_\mI(t,\bx)=e^{\im H_0t}\,\chi(t,\bx)
\end{equation}
and the time-dependent Schr\ödinger equation becomes
\begin{equation}
  \label{SEIP}
  \im\Partder{t}\,\chi_\mI(t,\bx)=H'_\mI(t)\,\chi_\mI(t,\bx)
\end{equation}
The \it{time-evolution operator}, defined by
\begin{equation}
  \label{EvolOp}
  \chi_\mI(t,\bx)=U_\mI(t,t_0)\,\chi_\mI(t_0,\bx)
\end{equation}
then satisfies the equation
\begin{equation}
  \im\Partder{t}\,U_\mI(t,t_0)=H'_\mI(t)\,U_\mI(t,t_0)
\end{equation}
with the solution~\cite[Eq. 6.23]{FW71}
\begin{equation}
  \label{Uexp}
  U_\mI(t,t_0)=1+\sum_{n=1}^\infty\frac{(-\im)^n}{n!}\int_{t_0}^t\dif^4x_n\cdots
  \int_{t_0}^t\dif^4x_1\;\TD\big[\calH'_\mI(x_n)\calH'_\mI(x_{n-1})
  \cdots\calH'_\mI(x_1)\big]
\end{equation}
Here, $x=(t,\bx)$, $\TD$ is the Dyson time-ordering operator, and
$\calH'_\mI(x)$ is the perturbation density defined by
\begin{equation}
  \label{calH}
  H'_\mI(t)=\intbx\,\calH'_\mI(t,\bx)
\end{equation}

In applying this formalism to perturbation theory, an
\it{adiabatic damping} is added~\cite{FW71}
\begin{equation}
  \label{Damp}
  H'_\mI(t)\rarr H'_{\mI\gamma}=H'_\mI\,e^{-\gamma\abs{t}}\,;\qquad
  U_\mI(t,t_0)\rarr U_{\mI\gamma}(t,t_0)
\end{equation}
where $\gamma$ is a small, positive number. This implies that as
$t\rarr-\infty$ the eigenfunctions of $H$ tend to eigenfunctions
of $H_0$.

In QED the perturbation density due to the interaction between the
electrons and the photon field is given by~\cite{Sch61}
\begin{equation}
  \label{Pert}
  \calH'_\mI(x)=-e\hpsi_\mI\dagg(x)\alpha^\mu
  A_\mu(x)\hpsi_\mI(x)
\end{equation}
where $e$ is the absolute value of the electronic charge,
$\hpsi_\mI\dagg(x),\,\hpsi_\mI(x)$ are the electron-field
operators in the interaction picture, $A_\mu$ the photon-field
operator and $\alpha^\mu$ are related to the standard Dirac
matrices by $\alpha^\mu=(1,\bs{\alpha})$.

\section{Green's function approach} \label{sec:Green} In this
section we shall essentially reproduce the derivation of the BS
equation by Gell-Mann and Low, starting from Green's functions. We
consider a two-particle system for which the Green's function is
defined~\cite[pp. 64 and 116]{FW71}
\begin{equation}
  \label{GFH}
  G(x'_1,x'_2;x_{10},x_{20})=
  -\frac{\bigbra{0_\mH}T_\mD[\psiH(x'_1)\psiH(x'_2)\psiH\dagg(x_{20})\psiH\dagg(x_{10})]\bigket{0_\mH}}
  {\langle0_\mH\ket{0_\mH}}
\end{equation}
Here, $\ket{0_\mH}$ represents the vacuum state and
$\psiH\dagg(x),\,\psiH(x)$ the electron-field operators, all in
the \it{Heisenberg representation}. The latter are related to
those in the interaction picture by
\begin{equation}
  \label{HP}
  \psiH(t,\bx)=U(0,t)\,\psiI(t,\bx)\,U(t,0)
\end{equation}
where $U$ is the evolution operator \eq{Uexp}. Transforming the
Green's function to the interaction picture then yields~\cite[Eq.
8.9]{FW71}, \cite[Eq. 16]{GML51}, \cite[Eq. 259]{LSA04}
\begin{equation}
  \label{GFI}
  G(x'_1,x'_2;x_{10},x_{20})=-\frac{\bigbra{0_\mI}T_\mD[\psiI(x'_1)\psiI(x'_2)
  U_\mI(\infty,-\infty)\psiI\dagg(x_{20})\psiI\dagg(x_{10})]\bigket{0_\mI}}
  {{\bra{0_\mI}U_\mI(\infty,-\infty)\ket{0_\mI}}}
\end{equation}
Obviously, only fully contracted terms contribute to the vacuum
expectation value. By applying \it{Wick's theorem}~\cite[p.
83]{FW71}~\cite[Sect. 11.5]{LM86}, this can be represented in
terms of \it{Feynman diagrams}. The denominator has the effect of
eliminating the singularities of the numerator, in the Feynman
picture represented by unlinked or disconnected diagrams, leading
to~\cite[Eq. 9.5]{FW71}
\begin{equation}
  \label{GFC}
  G(x'_1,x'_2;x_{10},x_{20})=-\bigbra{0_\mI}T_\mD[\psiI(x'_1)\psiI(x'_2)
  U_\mI(\infty,-\infty)\psiI\dagg(x_{20})\psiI\dagg(x_{10})\bigket{0_\mI}_\conn
\end{equation}

In contrast to the evolution operator \eq{EvolOp}, the Green's
function is relativistically \it{covariant} in the sense that the
integrations are performed over all space and time and the
electron-field operators can represent particle (positive-energy)
as well as hole (negative-energy) states. This also implies that,
in the energy representation (fourier transform), the energy is
conserved at all diagram vertices.

\begin{figure}[htb]
\begin{center}
\begin{picture}(5,3)(-1,0)
 \put(0,1){\circle*{0.15}}
 \put(2,1){\circle*{0.15}}
 \put(0,1){\Rect{2}{1}{\large\textbf{G}}}
 \put(0,1){\multiput(0,0)(0,0.1){10}{\line(1,0){2}}}
  \put(0,2){\circle*{0.15}}
 \put(2,2){\circle*{0.15}}
 \end{picture}
 \begin{picture}(4,3)(-1,0)
  \put(-1.75,1.5){\makebox(0,0){\Large =}}
  \put(-0.5,3){\makebox(0,0){$x'_1$}}
  \put(2.6,3){\makebox(0,0){$x'_2$}}
  \put(0,3){\circle*{0.15}}
 \put(2,3){\circle*{0.15}}
 \put(0,0){\Elline{3}{1.5}{}{}}
 \put(2,0){\Elline{3}{1.5}{}{}}
  \put(1,1.5){\makebox(0,0){$G_0$}}
  \put(0,0){\circle*{0.15}}
 \put(2,0){\circle*{0.15}}
  \put(-0.5,0){\makebox(0,0){$x_{10}$}}
  \put(2.6,0){\makebox(0,0){$x_{20}$}}
 \end{picture}
 \begin{picture}(4.5,3)(-2,0)
  \put(-1.5,1.5){\makebox(0,0){\Large +}}
  \put(0,3){\circle*{0.15}}
 \put(2,3){\circle*{0.15}}
  \put(0,1.5){\Elline{1.5}{0.75}{}{}}
 \put(2,1.5){\Elline{1.5}{0.75}{}{}}
 \put(0,1.4){\LineS{2}}
  \put(0,1.5){\LineS{2}}
 \put(0,1.6){\LineS{2}}
  \put(2.5,1.5){\makebox(0,0){$\cal K$}}
 \put(0,0){\Elline{1.5}{0.75}{}{}}
 \put(2,0){\Elline{1.5}{0.75}{}{}}
 \put(0,0){\circle*{0.15}}
 \put(2,0){\circle*{0.15}}
 \end{picture}
  \vsp\caption{Graphical representation of the two-particle Green's function
   \eq{GFK}. $\cal K$ represents \it{all} interactions between the electrons.}
   \label{Fig:GF}
 \end{center}
 \end{figure}
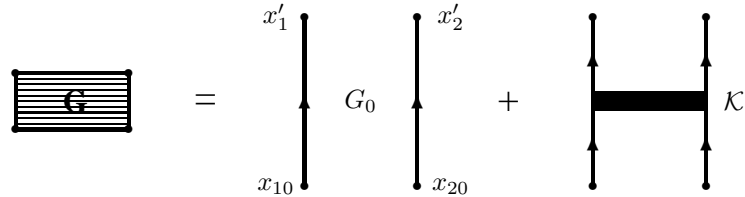

The Green's function can be expressed
\begin{eqnarray}
  \label{GFKx}
   &&G(x'_1,x'_2;x_{10},x_{20})=G_0(x'_1,x'_2;x_{10},x_{20})+\nn
  &&\ddint\dif^4 x_1\,\dif^4 x_2\,\dif^4 x_3\,\dif^4\,x_4\;
  G_0(x'_1,x'_2;x_1,x_2)\,{\cal K}(x_1,x_2;x_3,x_4)\,
  G_0(x_3,x_4;x_{10},x_{20})
\end{eqnarray}
where ${\cal K}$ represents the interaction kernel of all
connected diagrams and $G_0$ is the zeroth-order Green's function
\begin{eqnarray}
  \label{GF0}
  \Hsp G_0(x'_1,x'_2;x_{10},x_{20})&=&-\bigbra{0_\mI}T_\mD[\psiI(x'_1)\psiI(x'_2)
  \psiI\dagg(x_{20})\psiI\dagg(x_{10})]\bigket{0_\mI}\nn
  &=&\SF(x'_1,x_{10})\SF(x'_2,x_{20})
\end{eqnarray}
with $\SF$ being the Feynman \it{electron propagator} or
zeroth-order single-electron Green's function, defined by
\begin{equation}
  \label{SF}
  \im\SF(x',x_{0})=\bigbra{0_\mI}T_\mD[\psiI(x')\psiI\dagg(x_{0})]\bigket{0_\mI}
\end{equation}
assuming the vacuum state be normalized. This is illustrated in
Fig. \ref{Fig:GF}. In operator form the Green's function can be
expressed
\begin{equation}
  \label{GFK}
  G=G_0+G_0{\cal K} G_0
\end{equation}

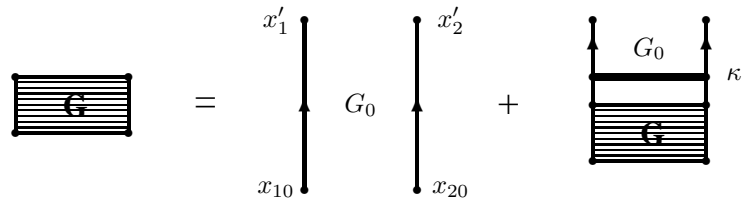
\begin{figure}[htb]
\begin{center}
\begin{picture}(5,3)(-1,0)
  \put(0,1){\circle*{0.15}}
 \put(2,1){\circle*{0.15}}
 \put(0,1){\Rect{2}{1}{\large\textbf{G}}}
  \put(0,1){\multiput(0.,0)(0,0.1){10}{\line(1,0){2}}}
  \put(0,2){\circle*{0.15}}
 \put(2,2){\circle*{0.15}}
 \end{picture}
 \begin{picture}(4,3)(-1,0)
  \put(-1.75,1.5){\makebox(0,0){\Large =}}
  \put(-0.5,3){\makebox(0,0){$x'_1$}}
  \put(2.6,3){\makebox(0,0){$x'_2$}}
  \put(0,3){\circle*{0.15}}
 \put(2,3){\circle*{0.15}}
 \put(0,0){\Elline{3}{1.5}{}{}}
 \put(2,0){\Elline{3}{1.5}{}{}}
  \put(1,1.5){\makebox(0,0){$G_0$}}
  \put(0,0){\circle*{0.15}}
 \put(2,0){\circle*{0.15}}
  \put(-0.5,0){\makebox(0,0){$x_{10}$}}
  \put(2.6,0){\makebox(0,0){$x_{20}$}}
 \end{picture}
 \begin{picture}(4.5,3)(-2,0)
  \put(-1.5,1.5){\makebox(0,0){\Large +}}
 \put(0,3){\circle*{0.15}}
 \put(2,3){\circle*{0.15}}
  \put(0,1.5){\Elline{1.5}{1.1}{}{}}
 \put(2,1.5){\Elline{1.5}{1.1}{}{}}
  \put(1,2.5){\makebox(0,0){$G_0$}}
  \put(0,1.97){\LineH{2}}
  \put(0,2.03){\LineH{2}}
 \put(0,2.0){\circle*{0.15}}
 \put(2,2.0){\circle*{0.15}}
 \put(2.5,2.05){\makebox(0,0){$\calKb$}}
 \put(0,1.5){\circle*{0.15}}
 \put(2,1.5){\circle*{0.15}}
  \put(0,0.5){\Rect{2}{1}{\large\textbf{G}}}
  \put(0,0.5){\multiput(0,0)(0,0.1){10}{\line(1,0){2}}}
 \put(0,0.5){\circle*{0.15}}
 \put(2,0.5){\circle*{0.15}}
 \end{picture}
  \vsp\caption{Graphical representation of the Dyson equation
   \eq{GF} for the two-particle Green's function. $\calKb$ represents
   the \it{non-separable} interactions between the electons.} \label{Fig:Dyson}
 \end{center}
 \end{figure}

In some cases the kernel of the Green's function can be separated
into two kernels
\begin{equation}
  \label{Ksep}
  {\cal K}={\cal K}_2G_0{\cal K}_1
\end{equation}
with no photon-field contractions between them. The kernel is then
said to be \it{separable}. If a kernel cannot be separated further
in this way, it is said to be \it{non-separable} \footnote{What we
here refer to as "separable" and non-separable" are often referred
to as "reducible" and "irreducible". Since the latter terms have
recently been used also with a different interpretation, we avoid
them here.}. The complete kernel can then be expressed
\begin{equation}
  \label{K}
  {\cal K}=\calKb+\calKb G_0\calKb+\calKb G_0\calKb
  G_0\calKb+\cdots
\end{equation}
where $\calKb$ represents all \it{non-separable} kernels. This
leads to the \it{Dyson equation} for the Green's function
\begin{equation}
  \label{GF}
  G=G_0+G_0\calKb G
\end{equation}
illustrated in Fig. \ref{Fig:Dyson}.

Bethe and Salpeter as well as Gell-Mann and Low argue that a
related equation can be set up for the two-electron bound-state
wave function. In that case the first (inhomogeneous) term on the
rhs does not contribute, since that is in their formulation
composed of \it{free-electron} propagators, and the bound-state
wave function does not have any such components. This leads to the
\it{homogeneous} equation
\begin{equation}
  \label{BS2}
   \Psi(x'_1,x'_2)=\ddint\dif^4 x_1\,\dif^4 x_2\,\dif^4
   x_3\,\dif^4x_4 \;G_0(x'_1,x'_2;x_1,x_2)\,\calKb(x_1,x_2;x_{3},x_{4})\,
  \Psi(x_{3},x_{4})\Hsp
\end{equation}
or in short-hand notations
\begin{equation}
  \label{BS2a}
   \Psi=G_0\,\calKb\,\Psi
\end{equation}
This is the original form of the Bethe-Salpeter equation~\cite[Eq.
11a]{SB51},~\cite[Eq. 37]{GML51}. It should be noted that this
wave function contains \it{individual times for the two
particles}. This reflects one of the problems referred to in the
Introduction. The relative time between the particles does not
correspond to any physical quantity and leads to spurious
solutions. There are several ways of eliminating the extra time
dependence in a covariant way. Sucher~\cite{Su58}, following
Salpeter~\cite{Salp52}, integrates the fourier transform over the
relative energy, which leads to a Schr\ödinger-like form with a
single time/energy dependence of the type \eq{BS} given above.
This reduction can be done without loosing any physical content of
the original equation~\cite{Conn91,PhilWal96,PTjon98,Bijt01}. In
the following sections we shall derive an equivalent equation in a
different way.

Our notations here differ from those used by Bethe-Salpeter and
Gell-Mann--Low. The Green's function \eq{GFH} is in their works
denoted by $K(12,34)$ and referred to as the "\it{amplitude
function for the propagation of the particles}" by
Bethe-Salpeter~\cite{SB51} and as the "\it{two-body kernel}" by
Gell-Mann--Low~\cite[Eq. 11]{GML51}. Our "non-separable kernel"
$\calKb$ is by BS denoted by $\bar{G}$ and referred to as
"\it{irreducible graphs}" and by GML denote by $G$ and referred to
as the "\it{interaction function}".

\section{Covariant evolution operator approach}
\label{sec:EvolOp}
 \subsection{Definitions}
In the following sections we shall derive the Bethe-Salpeter
equation, starting from the covariant form of the evolution
operator~\cite{LSA04}. This will demonstrate the relation between
the BS equation and standard many-body perturbation theory (MBPT)
in a clear way. In the present section we shall first review the
basics of the evolution-operator method and in the next section
use that method for deriving the BS equation. This will directly
lead to the Schr\ödinger-like form \eq{BS}.

According to the Gell-Mann--Low theorem~\cite[p. 61]{GML51,FW71}
the time-independent wave function \eq{wft} can in the case of a
single target function be expressed in intermediate normalization
(IN) \eq{IN} as
\begin{equation}
  \label{GML0}
  \bs {\Psi}(\bx)=\chi(0,\bx)=
  \limgam
  \frac{\Ugam{0}\Psi_0(\bx)}{\bra{\Psi_0}\Ugam{0}\ket{\Psi_0}}
\end{equation}
where $U_\gamma$ is the evolution operator \eq{Damp} and
$\Psi_0(\bx)$ is the time-independent zeroth-order wave function
\eq{ZOWF}. (From now on we work in the interaction picture and
leave out the subscript "I".) $\bs {\Psi}(x)$ is an eigenfunction
of the Hamiltonian $H_0+H'$
\begin{equation}
  \label{GML1}
  (H_0+H')\,\bs {\Psi}(\bx)=E\,\bs {\Psi}(\bx)
\end{equation}
where $H'$ is in our case the electron-field interaction
\eq{Pert}. Since this perturbation represents an uncontracted
photon, the wave function $\bs {\Psi}(\bx)$ will generally lie in
an \it{extended Fock space}, where the number of photons is not
conserved.

The GML formula can be generalized to a general multi-dimensional
model space~\cite[Eq. 110]{LSA04}
\begin{equation}
  \label{GML}
  \bs {\Psi}^\alpha(\bx)=\limgam \frac{N^\alpha\Ugam{0}\Phi^\alpha(\bx)}
  {\bra{\Phi^\alpha}\Ugam{0}\ket{\Phi^\alpha}}\,;\qquad
  (\alpha=1,2,\cdots d)
\end{equation}
where the function $\Phi^\alpha$ is defined
\begin{equation}
  \label{Phi}
  \Phi^\alpha(\bx)=\limgam\lim_{t\rarr-\infty}\chi^\alpha(t,\bx)
\end{equation}
This function is generally distinct from the zeroth-order wave
function \eq{ZOWF} in intermediate normalization. Since the
function \eq{Phi} generally does not satisfy IN, a normalization
constant $N^\alpha$ is inserted.

\begin{figure}[htb]
\begin{center}
 \begin{picture}(6,3.5)(-5,0)
 \put(-5.5,1.5){\makebox(0,0){\Large$U_{\rm{Noncov}}(t',t_0)=$\;\;$1\;\;+$}}
 \put(-0.75,3){\multiput(0.05,0)(0.25,0){14}{\line(1,0){0.15}}}
 \put(-1.5,3){\makebox(0,0){t\,=\,t'}}
   \put(0,1.5){\Elline{1.5}{0.75}{\hpsi_+\dagg\;}{}}
 \put(2,1.5){\Elline{1.5}{0.75}{}{\;\hpsi_+\dagg}}
 \put(0,1.4){\LineS{2}}
 \put(0,1.5){\LineS{2}}
 \put(0,1.6){\LineS{2}}
 \put(2.5,1.5){\makebox(0,0){$\cal K$}}
 \put(0,0){\Elline{1.5}{0.75}{\hpsi_+\;}{}}
 \put(2,0){\Elline{1.5}{0.75}{}{\;\hpsi_+}}
  \put(-0.75,0){\multiput(0.05,0)(0.25,0){14}{\line(1,0){0.15}}}
 \put(-1.5,0){\makebox(0,0){t\,=\,$t_0$}}
 \end{picture}
  \vsp\caption{Graphical representation of the non-covariant evolution
 operator \eq{NoncovEv}. The time evolution occurs only in the positive direction.}
 \label{Fig:Noncov}
 \end{center}
 \end{figure}

\begin{figure}[htb]
\begin{center}
 \begin{picture}(6,3.5)(-5,-0.5)
 \put(-5,1){\makebox(0,0){\Large$U_{\rm{Cov}}(t',t_0)=$\;\;$1\;\;+$}}
 \put(-0.75,2){\multiput(0.05,0)(0.25,0){14}{\line(1,0){0.15}}}
 \put(-1.5,2){\makebox(0,0){t\,=\,t'}}
 \put(0,2){\circle*{0.15}}
 \put(2,2){\circle*{0.15}}
 \put(0,2){\Elline{1}{0.5}{\psi\dagg}{}}
 \put(2,2){\Elline{1}{0.5}{}{\,\,\psi\dagg}}
 \put(0,1){\Elline{1}{0.55}{}{}}
 \put(2,1){\Elline{1}{0.55}{}{}}
 \put(0,0.9){\LineS{2}}
 \put(0,1){\LineS{2}}
 \put(0,1.1){\LineS{2}}
 \put(1,1.5){\makebox(0,0){$G_0$}}
 \put(2.5,1){\makebox(0,0){$\cal K$}}
 \put(1,0.4){\makebox(0,0){$G_0$}}
 \put(0,0){\Elline{1}{0.45}{}{}}
 \put(2,0){\Elline{1}{0.45}{}{}}
 \put(0,-1){\Elline{1}{0.5}{\psi}{}}
 \put(2,-1){\Elline{1}{0.5}{}{\psi}}
 \put(0,0){\circle*{0.15}}
 \put(2,0){\circle*{0.15}}
 \put(-0.75,0){\multiput(0.05,0)(0.25,0){14}{\line(1,0){0.15}}}
 \put(-1.5,0){\makebox(0,0){t\,=\,$t_0$}}
 \end{picture}
  \begin{picture}(6,3.5)(-6,-0.5)
  \put(-3,1){\makebox(0,0){\Large=}}
 \put(-0.75,1.5){\multiput(0.05,0)(0.25,0){14}{\line(1,0){0.15}}}
 \put(-1.5,1.5){\makebox(0,0){t\,=\,t'}}
 \put(0,1.5){\Elline{1}{0.5}{\psi\dagg}{}}
 \put(2,1.5){\Elline{1}{0.5}{}{\,\,\psi\dagg}}
  \put(0,1.5){\circle*{0.15}}
 \put(2,1.5){\circle*{0.15}}
 \put(0,0.5){\Rect{2}{1}{\large\textbf{G}}}
  \put(0,0.5){\multiput(0,0)(0,0.1){10}{\line(1,0){2}}}
  \put(0,1.5){\circle*{0.15}}
 \put(2,1.5){\circle*{0.15}}
 \put(0,-.5){\Elline{1}{0.5}{\psi}{}}
 \put(2,-.5){\Elline{1}{0.5}{}{\psi}}
 \put(0,0.5){\circle*{0.15}}
 \put(2,0.5){\circle*{0.15}}
 \put(-0.75,0.5){\multiput(0.05,0)(0.25,0){14}{\line(1,0){0.15}}}
 \put(-1.5,0.5){\makebox(0,0){t\,=\,$t_0$}}
 \end{picture}
 \vsp\caption{Graphical representation of the covariant evolution
 \eq{CovEv} (left). Here, time evolution can occur in the
 positive as well as the negative direction. The right part of the figure
 depicts the relation to the two-times Green's function \eq{CovGF}.}
 \label{Fig:CovGF}
 \end{center}
 \end{figure}

For a two-electron system the \it{non-covariant} evolution
operator \eq{EvolOp} can in analogy with the Green's function
\eq{GFK} be expressed
\begin{equation}
  \label{NoncovEv}
  U_{\rm{Noncov}}(t',t_0)=1+\hpsi_+\dagg(x'_1)\hpsi_+\dagg(x'_2)\,
  {\cal K}\,\hpsi_+(x_{20})\hpsi_+(x_{10})
\end{equation}
where again ${\cal K}$ represents the kernel of all fully
contracted (separable and non-separable) interactions and
$\hpsi_+\dagg,\,\hpsi_+$ the positive-energy part of the
electron-field operators. This is illustrated in Fig.
\ref{Fig:Noncov}. In contrast to the Green's function above, the
evolution operator \eq{Uexp} has a single initial time $t=t_0$ and
a single final time $t=t'$. The time integration is performed from
$t=t_0$ to $t=t'$
--- only in the positive direction --- which implies that the
operator is \it{not relativistically covariant}.

A fully covariant form of the evolution operator that is
applicable to relativistic problems can be obtained by inserting
electron propagators in the non-covariant expression, as indicated
in Fig. \ref{Fig:CovGF} (left), corresponding to the
expression~\cite[Sect. 5]{LAS01,LSA04}
\begin{equation}
  \label{CovEv}
  U_\Cov(t',t_0)=1+
  \ddint\dif^3\bx'_1\,\dif^3\bx'_2\,\dif^3\bx_{10}\,\dif^3\bx_{20}\;
  \hpsi\dagg(x'_1)\hpsi\dagg(x'_2)\,G_0{\cal
  K}G_0\,\hpsi(x_{20})\hpsi(x_{10})
\end{equation}
leaving out the integrations over the coordinates of ${\cal
  K}$ (see Eq. \ref{GFKx}). It then
follows from the relation \eq{GFK} that the covariant evolution
operator is related to the \it{two-times} Green's function (where
all initial and all final times are equal) by
\begin{equation}
  \label{CovGF}
  U_\Cov(t',t_0)=\ddint\dif^3\bx'_1\,\dif^3\bx'_2\,\dif^3\bx_{10}\,\dif^3\bx_{20}\;
  \hpsi\dagg(x'_1)\hpsi\dagg(x'_2)\,G(x_1',x_2';x_{10},x_{20})
  \,\hpsi(x_{20})\hpsi(x_{10})
\end{equation}
as illustrated in Fig. \ref{Fig:CovGF} (right).

From the relations~\cite[Eq. 193 (note misprints)]{LSA04}
\begin{eqnarray}
  \label{Rel}
  \Hsp\intbx_0\, \im\SF(x,x_0)\,\hpsi(x_0)=\Theta(t-t_0)\,\hpsi_+(x)-
  \Theta(t_0-t)\,\hpsi_-(x)\nn
  \Hsp\intbx\, \hpsi\dagg(x)\,\im\SF(x,x_0)=\Theta(t-t_0)\,\hpsi_+\dagg(x_0)-
  \Theta(t_0-t)\,\hpsi_-\dagg(x_0)
\end{eqnarray}
it follows directly that the form \eq{CovEv} is equivalent to the
non-covariant form \eq{NoncovEv}, when only particle states are
involved. That the former in addition is relativistically
covariant follows from the fact that \it{the electron-field
operators can represent particle as well as hole states and the
internal time integrations are performed over all times}--- in the
positive as well as the negative direction. From now on we shall
work only with the covariant form of the evolution operator and
leave out the subscript $"_\Cov$".

In using the evolution operator in perturbation theory, we assume
that we operate to the far right on positive-energy states in the
model space. Then, as shown in Appendix \ref{App:SingPhot}, we can
eliminate the rightmost zeroth-order Green's function and set the
initial time to $t_0=-\infty$. We shall also assume that the limit
of the adiabatic damping $\gamma\rarr0$ is taken.

The Covariant evolution operator is closely related to the Green's
function---the main difference being that the Green's function is
a \it{function}, while the evolution operator is an \it{operator}.
The poles of the Green's function (in the energy representation)
correspond to the energies of the system, while it gives no direct
information about the wave function. The covariant evolution
operator, on the other hand, contains information about the energy
as well as the wave function.

 \subsection{Model-space contributions}
 \label{sec:MSC}
                                       \renewcommand{\Ugam}[1]{U(#1,-\infty)}
                                       \renewcommand{\Ugamtil}[1]{\widetilde{U}(#1,-\infty)}
                                       \renewcommand{\Ugamt}{\widetilde{U}}
                                       \newcommand{\cdd}{\cdot\cdot}
                                        \renewcommand{\Ugam}[1]{\widetilde{U}(#1)}
                                         \renewcommand{\Ugamtil}[1]{\widetilde{U}(#1)}
                                                  \newcommand{\ett}{^{(1)}}
                                                  \newcommand{\tva}{^{(2)}}
                                                  \newcommand{\tre}{^{(3)}}
                                                  \newcommand{\fyr}{^{(4)}}
                                                  \newcommand{\MSC}{\rm{MSC}}
                                                  \newcommand{\Htil}{\widetilde{H}}
                                                  \newcommand{\Ubar}{\bar{U}}
                                                  \newcommand{\Hbar}{\bar{H}}
                                                  \newcommand{\Udot}{\dot{U}}
                                                  \newcommand{\Ubardot}{\dot{\Ubar}}
                                                  \newcommand{\Utildot}{\dot{\Util}}
                                                  \newcommand{\Cdot}{\dot{C}}
                                                  \newcommand{\str}{\hspace{-2.5mm}/}
                                                        \newcommand{\E}{{\calE}}

Even after eliminating unlinked or disconnected contributions in
Eq. \eq{GFC}, the evolution operator may contain
(quasi)singularities, namely when the intermediate state of a
separable kernel lies in the model space and is degenerate or
nearly degenerate (quasi-degenerate) with the initial state. As
mentioned, a kernel is said to be \it{separable}, if it can be
separated into two kernels with no photon contractions between
them. Singularities appear only for separable interactions. In the
covariant-evolution-operator approach these singularities are
eliminated by introducing a \it{reduced evolution operator}
$\Util(t,-\infty)$~\cite[Eq. 116]{LAS01,LSA04}, defined by
\begin{equation}
  \label{Ured0}
  U(t,-\infty)P=P+\Util(t,-\infty)P\bsdot PU(0,-\infty)P
\end{equation}
Here, the last term is a product of two operators that evolve
\it{independently} from an initial state in the model space
($t=-\infty$), which is indicated by the "dot". Note also that the
last factor has the final time $t=0$ and hence is time
independent. This situation should be distinguished from the case
where two operators are "coupled" and operate "in succession"
\begin{equation}
  \label{Usuc}
  U(t,t_0)=U(t,t")\,U(t",t_0)
\end{equation}
This distinction will be important for the following treatment.

Normally, we shall assume that the initial time in the evolution
operator is $t_0=-\infty$, and in cases where there is no risk for
ambiguity we shall leave that out from the operator, so that
\[U(t)={U(t,-\infty)}\]
The definition \eq{Ured0} will then be written
\begin{equation}
  \label{Ured}
  \boxed{U(t)P=P+\Util(t)P\bsdot PU(0)P}
\end{equation}
We also introduce the notation $U'(t)=U(t)-1$, which yields in
place of the definition \eq{Ured}
\begin{equation}
  \label{Ured2}
  \Util(t)P=U'(t)P-\Util(t)P\bsdot PU'(0)P
\end{equation}
Here, the last term is the \it{counterterm}
\begin{equation}
  \label{C}
  \boxed{C(t)P=-\Util(t)P\bsdot PU'(0)P}
\end{equation}
which removes the (quasi)singularities. This can also be expressed
\begin{equation}
  \label{Cexp}
  C(t)P=-\Util(t)P\bsdot P\Util(0)P
  -\Util(t)P\bsdot P\Util(0)P\bsdot P\Util(0)P-\cdots
\end{equation}

After removing a singularity, there is normally a non-vanishing
remainder, referred to as the \it{model-space contribution} (MSC),
defined as
\begin{equation}
  \label{Mdef}
  \boxed{MP=\Util(0)P-\Ubar(0)P}
\end{equation}
and further discussed in the Appendices. The new operator $\Ubar$
(\it{"U-bar"}) is defined as the evolution operator with \it{all
model-space states removed}. (The MSC is analogous to the
\it{reference-state contribution}, appearing in the $S$-matrix
formalism, where the effect normally appears only when the
intermediate states is equal to the reference or initial state. In
our formalism with an extended model space the effect can appear
also for other model-space states, and we prefer the more general
term.) It should be noted that the counterterms also remove
quasi-singularities, due to quasi-degenerate states that are
included in the model space. This can be of vital importance for
the convergence of the procedure.

As discussed in Appendix \ref{App:SepPh}, the model-space
contributions are of \it{two kinds}. The first kind appears for
all interactions, even if they are time or energy independent,
while the second kind appears only for time- or energy-dependent
interactions. The first kind appears also in standard
time-independent perturbation theory and corresponds to so-called
\it{folded diagrams} of MBPT~\cite[Fig. 5]{LM86,LSA04}.

\subsection{The wave operator and effective interaction}
\label{sec:Wop} As mentioned previously, the evolution operator
\eq{Uexp} with the perturbation \eq{Pert} can contain uncontracted
photon operators, which implies that it operates in a general
\it{Fock space,} where the number of virtual photons is not
conserved. We then separate the covariant evolution operator
\eq{CovEv} into
\begin{equation}
  U(t)=PU(t)+\Q U(t)
\end{equation}
where $\Q=1-P$ is operating in the general Fock space, while $P$
is the projection operator for the model space, confined to the
restricted Hilbert space with no uncontracted photon. This leads
with the definition \eq{Ured} of the reduced evolution operator
for $t=0$ to the \it{factorization theorem}~\cite[Eq. 121]{LSA04}
\begin{equation}
  \label{Fact}
  \boxed{U(0)P=\big[1+\Q\Util(0)\big]P\bsdot PU(0)P}
\end{equation}
where the first factor on rhs is regular. Inserted in the GML
formula \eq{GML}, this yields
\begin{equation}
  \bs{\Psi}^\alpha(\bx)=\big[1+\Q\Util(0)\big]\,\Psi_0^\alpha
\end{equation}
where $\Psi_0^\alpha$ is the zeroth-order wave function (ZOWF)
\eq{ZOWF} in intermediate normalization
\begin{equation}
  \Psi^\alpha_0(\bx)=P\bs{\Psi}^\alpha(\bx)=
  \frac{N^\alpha
  PU(0)\Phi^\alpha(\bx)}{\bra{\Phi^\alpha}U(0)\ket{\Phi^\alpha}}
\end{equation}
The square bracket above is the \it{wave operator}
\begin{eqnarray}
  \label{WaveEqn}
  \bs{\Om}&=&1+\Q\Util(0)\nn
  \hspace{5cm}\bs{\Psi}^\alpha(\bx)&=&\bs{\Om}\,\Psi_0^\alpha(\bx)
\end{eqnarray}
The result here is a direct consequence of the generalized
Gell-Mann--Low theorem and the definition of the reduced evolution
operator.

As mentioned, with the perturbation \eq{Pert} the wave function
$\bs{\Psi}^\alpha$ lies generally in a Fock space where the number
of (virtual) photons is not conserved. But we are interested here
in the case where all photon operators are fully contracted, and
for that purpose we project the equation on the restricted Hilbert
space without uncontracted photon operators
\begin{equation}
  {\cal P}\bs{\Psi}^\alpha(\bx)={\cal P}\big[1+\Q\Util(0)\big]\,\Psi_0^\alpha(x)
\end{equation}
or
\begin{equation}
  \label{wf}
  \Psi^\alpha(\bx)=\big[1+Q\Ugamtil{0}\big]\Psi^\alpha_0(\bx)
\end{equation}
where $\Psi^\alpha(\bx)={\cal P}\bs{\Psi}^\alpha(\bx)$ is the
projected wave function on the restricted Hilbert space and
$Q={\cal P}\Q$ is the conventional projection operator for the
complementary space (outside the model space). The wave operator
in this space is
\begin{equation}
  \label{WaveOp}
  \boxed{\Omega={\cal P}\bs{\Om}=1+Q\Util(0)}
\end{equation}
In IN \eq{IN} the wave operators satisfy in both spaces the
relation \eq{POP}
\begin{equation}
  \label{OmIN}
  P\bs{\Om} P=P\Om P=P.
\end{equation}

The \it{effective interaction} \eq{EffInt2} is in this formalism
given by~\cite[Eq. 130]{LSA04}
\begin{equation}
  \label{EffInt}
  \boxed{H'\eff=P\Big[\im\Partder{t}\Ugamtil{t}\Big]_{t=0}P}
\end{equation}

\section{Connection to the Bethe-Salpeter equation}
\label{sec:Appl} \subsection{Expansion of the wave operator}

We know from the generalized Gell-Mann--Low relation \eq{GML} that
the wave function $\bs{\Psi}^\alpha(\bx)$ in the extended Fock
space satisfies a Schr\ödinger-like equation \eq{GML1} with the
Hamiltonian $H=H_0+H'$, where $H'$ is the perturbation \eq{Pert}.
We now want to find the corresponding equation for the wave
function $\Psi^\alpha={\cal P}\bs{\Psi}^\alpha(\bx)$ in the
restricted space with no uncontracted photons, and we shall see in
this section that this leads to the \it{Bethe-Salpeter equation}.

We shall start with the exchange of a sequence of separable
covariant single photons between the electrons, which can then be
generalized to other interactions, leading to the full equation.
This will first be done for a degenerate model space and then
extended to the general case.

As shown in Appendix \ref{App:SingPhot} (Eq. \ref{Om1}), the
contribution to the wave operator due the exchange of one single
photon is
\begin{equation}
  \label{Om11}
  \Omega\ett P=Q\Util\ett(0,\E)P= \GQ(\E) V(\E)P
  \end{equation}
and the corresponding contribution to the effective Hamiltonian
$H\eff\ett(\E)= P V(\E)P$. Here, $\GQ(\E)$ is the "reduced"
resolvent \eq{RedRes} and $V(\calE)$ is the effective
single-photon potential \eq{VExpl}, assuming that we operate to
the right on a fourier transform \eq{FE} with the energy parameter
$\calE$.

Similarly, it is demonstrated in Appendix \ref{App:SepPh} (Eq.
\ref{Om2}) that the contribution to the evolution operator from
\it{two} separable single-photon interactions is for a degenerate
model space given by (leaving out the arguments)
\begin{equation}
  \label{Om2alt}
  \Omega\tva P=Q\Util\tva P=\GQ V\Omega\ett
  P+\partder{\Omega\ett}{\E}*H\eff\ett
\end{equation}
where the last term represents the \it{model-space contribution}
(MSC) \eq{Mdef}
\begin{equation*}
  QM\tva P=\partder{\Omega\ett}{\E}*H\eff\ett
\end{equation*}
(The asterisk is introduced here only to indicate that there is a
cancelled singularity at that position, which is of importance for
the further treatment, as discussed in the Appendices.) The
contribution to the effective Hamiltonian \eq{Heff2} due to
two-photon exchange is
\begin{equation}
  \label{TwoPhEff}
  H\eff\tva= PV\Omega\ett P+\partder{H\eff\ett}{\E}*H\eff\ett
  =\Hbar\eff\tva+\partder{H\eff\ett}{\E}*H\eff\ett
\end{equation}
The last term is the MSC to the effective interaction, and if the
model space is degenerate with the energy $E_0$ that term becomes
\begin{equation}
  \label{HMSC}
  \partder{H\eff\ett}{\E}*H\eff\ett=P\partder{V(\E)}{\E}\Big|_{E_0}PV(E_0)P
\end{equation}
This corresponds to the "reference-state contribution", discussed
in connection with the $S$-matrix treatment of two-photon
exchange~\cite{BMJ93,LPS95}.

The treatment above will now be generalized to all orders as a
first step towards deriving the full BS equation. We start with
the covariant evolution operator \eq{CovEv} $U(t)=U(t,-\infty)$
and the reduced evolution operator \eq{Ured2}
\begin{equation}
  \label{UredA}
  \Util(t)P=U'(t)P-\Util(t)\bsdot PU'(0)P
\end{equation}
where $U'=U-1$. Note that only the first factor in the product is
time dependent (see. Eq. \ref{U2}). Note also the appearance of
the "dots" in this expression.  The significance of the dot is
discussed in relation to the definition \eq{Ured0}.

In the following we shall leave out the prime on $U'$ and also the
time arguments, if there is no risk of ambiguity. We then express
the counterterm \eq{C} as
\begin{equation}
  \label{CA}
  \boxed{CP=-\Util\bsdot PUP}
\end{equation}
and the evolution operator is given by
\begin{equation}
  \label{UA}
  UP=\G VP+\G V\G VP+\G V\G V\G VP+\cdots
\end{equation}
where $\G=\G(\E)$ is the resolvent \eq{Gamma1}. The "\it{U-bar}"
operator \eq{Mdef}, with all intermediate model-space states
removed, is
\begin{equation}
  \label{Ubar}
  \Ubar P=\G VP+\GQ V\GQ VP+\GQ V\GQ V\GQ VP+\cdots
\end{equation}

We introduce a special symbol for the time derivative at time
$t=0$
\begin{equation}
  \label{Adot}
  \boxed{\dot{A}=\im\partder{A}{t}\Big|_{t=0}}
\end{equation}
Since the evolution operator \eq{SPc} has the time dependence
\[U(t,\E)=e^{-\im t(\E-H_0)}\,U(0,\E)\]it follows that the time
derivation eliminates the denominator of the first (leftmost)
resolvent, so that
\begin{eqnarray}
  \label{dots}
  &&P\Udot P=P\big(V+V\G V+V\G V\G V+\cdots\big)P=PVP+PVUP\nn
  &&P\Ubardot P=P\big(V+V\GQ V+V\GQ V\GQ V+\cdots\big)P=PVP+PV\Ubar P
  \end{eqnarray}
The \it{effective interaction} $H'\eff$ \eq{EffInt} is with this
notation given by
\begin{equation}
  \label{Utildot}
  \boxed{H'\eff=P\Utildot P}
\end{equation}
We also introduce the corresponding  \it{"H-bar"} operator with no
intermediate model-space states
\begin{equation}
  \label{Hbar}
  \Hbar'\eff=P\Ubardot P
\end{equation}

We recall the definition \eq{Mdef} of the model-space contribution
(MSC)
\begin{equation}
  \label{MSCA}
  \Util P=\Ubar P+MP
\end{equation}
and can easily derive the identities
\begin{eqnarray}
  \label{ID}
  \Hsp UP&=&\Ubar P+\Ubar\;PUP=\Ubar P+\Util\;PUP-M\;PUP\\
  \label{ID2}
  \Hsp\Ubar P&=&UP-U\;PUP+U\;PUP\;PUP-\cdots
\end{eqnarray}
Then the reduced evolution operator \eq{UredA} becomes
\begin{equation}
  \label{UtilA}
  \Util P=\Ubar P+CP=\Ubar P+\Util\,PUP-\Util\bsdot PUP-M\,PUP
\end{equation}
which using the definition \eq{MSCA} leads to the series
\begin{equation}
    \label{OmRec}
  \Util P=\Ubar P+\big(\Util\; PUP-\Util\bsdot PUP\big)
  \big(1-PUP+ PUP\,  PUP+\cdots\big)\Hsp
\end{equation}
With the identity \eq{ID2} this becomes
\begin{equation}
  \label{OmRec2}
  \boxed{\Util P=\Ubar P+\big(\Util\; P\Ubar P-\Util\bsdot P\Ubar
  P\big)}
\end{equation}
which is an exact expression also for a quasi-degenerate model
space. It can be expanded as
\begin{equation}
  \label{OmRec3}
  \Util P=\Ubar P+\big(\Ubar\; P\Ubar P-\Ubar\bsdot P\Ubar P\big)
  +\big(\Ubar\; P\Ubar P-\Ubar\bsdot P\Ubar P\big)
  \big( P\Ubar P-\bsdot P\Ubar P\big)+\cdots
\end{equation}
As discussed in Appendix \ref{App:SepPh}, the result \eq{OmRec2}
can be expressed
\begin{equation}
  \label{OmRec2A}
  \Util P=\Ubar P+\partdelta{\Util}{\E}*P\Ubardot P
  =\Ubar P+\partdelta{\Util}{\E}*\Hbar'\eff
\end{equation}
where $\delta\E$ is the change in the model-space energy,
represented by the "dot", $\delta\Util$ is the corresponding
change in $\Util$, and $\Hbar'\eff$ is the "H-bar" operator
\eq{Hbar}. In the case of complete degeneracy this becomes
\begin{equation}
  \label{UTot}
  \Util P=\Ubar P+\partder{\Util}{\E}\Big|_{\E=E_0}*\Hbar'\eff
\end{equation}
Introducing the\it{"Omega-bar"} operator $\Ombar$ (with no
intermediate model-space states) in analogy with the wave operator
\eq{WaveOp}
\begin{equation}
  \label{Ombar}
  \Ombar P=P+Q\Ubar P=P+\GQ VP+\GQ V\GQ VP+\cdots
\end{equation}
we can express the relations above as
\begin{equation}
  \label{OmTot}
  \boxed{\Om P=\Ombar P+\partdelta{\Om}{\E}*\Hbar'\eff
  \Rarr\Ombar P+\partder{\Om}{\E}\Big|_{\E=E_0}*\Hbar'\eff}
\end{equation}
The second term is here consequently an exact expression for the
entire model-space contribution to the wave operator. This is in
agreement with the three-photon result \eq{Om3C}.

By taking the time derivative of the relation \eq{OmRec2}, using
the relations above, we obtain similarly
\begin{equation}
  \label{HHbar}
  \boxed{H'\eff=\Hbar'\eff+\partdelta{H'\eff}{\E}*\Hbar'\eff
  \Rarr\Hbar'\eff+\partder{H'\eff}{\E}\Big|_{\E=E_0}*\Hbar'\eff}
\end{equation}
The second term represents here the model-space contribution to
the effective interaction. This result agrees also with the
third-order result \eq{Heff3}.

From the results above we conjecture that the wave operator can at
complete degeneracy alternatively be expressed
\begin{equation}
  \label{Om}
  \Omega P=\Ombar P+\partder{\Ombar}{\E}*H'\eff
  +\half\ppartder{\Ombar}{\E}*\big(H'\eff\big)^2
  +\tref\pppartder{\Ombar}{\E}*\big(H'\eff\big)^3+\cdots
  =\Ombar P+\sum_{n=1}^\infty\npartder{n}{\Ombar}{\E}\;
  *\big(H'\eff\big)^{n}
\end{equation}
with all derivatives taken at $\E=E_0$, and we shall now prove
this relation by showing that it is compatible with the results
\eq{OmTot} and \eq{HHbar}, which we have rigorously derived. This
equation contains eliminated singularities, indicated by the
asterisks. As discussed in the Appendices, the derivative of such
an expression has to be taken \it{before} the singularity is
eliminated. Using the rules developed, particularly in Appendix
\ref{App:Exp}, we find for instance
\begin{eqnarray}
  \label{Deriv2}
  \Hsp\partder{}{\E}\Big(\partdelta{\Ombar}{\E}*H'\eff\Big)
  &\Rarr&\half\ppartder{\Ombar}{\E}*H'\eff
  +\partder{\Ombar}{\E}*\partder{H'\eff}{\E}\\
  \label{Deriv3}
  \Hsp\partder{}{\E}\Big(\half\ppartdelta{\Ombar}{\E}*\big(H'\eff\big)^2\Big)
  &\Rarr&\npartder{3}{\Ombar}{\E}*\big(H'\eff\big)^2
  +\half\ppartder{\Ombar}{\E}*\partder{H'\eff}{\E}*H'\eff
\end{eqnarray}
Note that in the second example the two $\Hbar'\eff$ operators
have in the quasi-degenerate case different energy parameters, and
therefore only one of them is affected by the derivation.

Generalizing these rules, we can evaluate the derivative of the
wave operator \eq{Om}
\begin{eqnarray}
  \label{OmDerA}
  \Hsp\partder{\Om}{\E}&=&\partder{\Ombar}{\E}
  +\half\ppartder{\Ombar}{\E}* H'\eff
  +\tref\pppartder{\Ombar}{\E}*\big(H'\eff\big)^2+\cdots\nn
  \Hsp&+&\partder{\Ombar}{\E}*\partder{H'\eff}{\E}
  +\half\ppartder{\Ombar}{\E}*\partder{H'\eff}{\E}*H'\eff
  +\tref\pppartder{\Ombar}{\E}*\partder{H'\eff}{\E}*(H'\eff)^2+\cdots
\end{eqnarray}
or
\begin{equation}
  \label{OmDerB}
  \partder{\Om}{\E}=\sum_{n=1}^\infty
  \npartder{n}{\Ombar}{\E}\;\Big[\big(H'\eff\big)^{n-1}
  +\partder{H'\eff}{\E}*(H'\eff)^{n-1}\Big]
\end{equation}
We now insert this expression into the equation \eq{OmTot}, which
yields
\begin{equation}
  \partder{\Om}{\E}*\Hbar'\eff=\sum_{n=1}^\infty
  \npartder{n}{\Ombar}{\E}\;\Big[\big(H'\eff\big)^{n-1}
  \Hbar'\eff+\partder{H'\eff}{\E}*(H'\eff)^{n-1}\Hbar'\eff\Big]
\end{equation}
or, using the relation \eq{HHbar},
\begin{equation}
  \label{Key}
  \boxed{\Om P=\Ombar P+\sum_{n=1}^\infty
 \npartder{n}{\Ombar}{\E}\;\big(H'\eff\big)^{n}}
\end{equation}
This is identical to the conjectured relation \eq{OmTot} and
therefore completes the proof. The sum represents by definition
the model-space contribution (MSC).

\begin{figure}[htb]
\begin{center}
\begin{picture}(5,4.5)(-2,0)
\put(0,0){\lline{4}{0.75}{}{}} \put(2,0){\lline{4}{0.75}{}{}}
\put(0,1){\Crossphotons{}{}{}{}{}{}}
\end{picture}
\begin{picture}(5,4.5)(-2,0)
\put(0,0){\lline{4}{0.75}{}{}} \put(2,0){\lline{4}{0.75}{}{}}
\put(0,0.5){\photonNE{}{}{}} \put(0,1.5){\photonNE{}{}{}}
\end{picture}
\begin{picture}(5,4.5)(-2,0)
\put(0,0){\lline{4}{0.75}{}{}} \put(2,0){\lline{4}{0.75}{}{}}
\put(0,2){\photon{}{}{}} \put(2,2){\ElSE{}{}{}}
\end{picture}
    \renewcommand{\normalsize}{\footnotesize}
    \caption{Examples of non-separable two-photon interactions.}
          \renewcommand{\normalsize}{\standard}
    \label{Fig:NonSep}
  \end{center}
\end{figure}
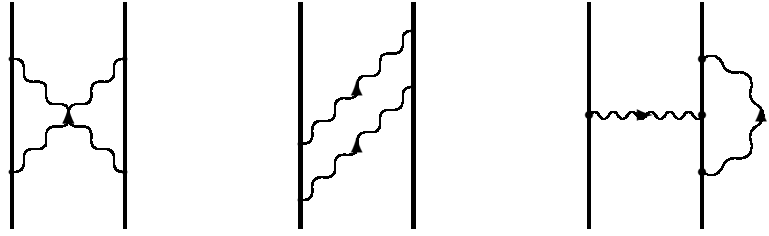

\subsection{Derivation of the Bethe-Salpeter equation. Degenerate model space.}

The previous treatment has been based upon the Hamiltonian
$H=H_0+V(E)$, where $V(E)$ is the potential due to the exchange of
a single covariant photon. But the process can be repeated in
exactly the same way, if we include \it{all non-separable
multi-photon interactions}. A non-separable interaction is defined
as an interaction that cannot be represented by two or more
simpler interactions in the way treated here. Two
photons---crossing or noncrossing--- that overlap in time
represent non-separable two-photon interactions (see Fig.
\ref{Fig:NonSep}, c.f. also Ref.~\cite[Fig. 1]{SB51}). These can
also include the radiative self-energy and vertex corrections. In
a similar way non-separable three-, four-,... photon interactions
can be defined. Therefore, in the following we replace the
single-photon potential $V$ by the general potential due to all
non-separable interactions
\begin{equation}
  \label{Vgen}
  \calV(E)=V(E)+V_2(E)+V_3(E)+\cdots
\end{equation}

As discussed in the Appendices, when operating on a fourier
transform of definite energy $\E$, the energy parameter of $\Ombar
P$ is equal to that energy, i.e.,
\begin{equation}
  \label{OmbarE}
  \Ombar F(\E)=\Ombar(\E)F(\E)
\end{equation}
For a degenerate model space of energy $E_0$ this means that
\begin{equation}
  \label{OmbarE0}
  \Ombar P=\Ombar(E_0)P=P+\GQ(E_0)\V(E_0)+\cdots
\end{equation}

The model functions are eigenfunctions of the effective
Hamiltonian \eq{SecEq}, and for a degenerate model space (of
energy $E_0$) they are eigenfunctions also of the effective
interaction \eq{EffInt2},
\begin{equation}
  \label{DeltaE}
  H'\eff\ket{\Psi_0^\alpha}=\Delta E^\alpha\ket{\Psi_0^\alpha}
  \end{equation}
where $\Delta E^\alpha=E^\alpha-E_0$. Operating with the operator
equation \eq{Key} directly on the model function $\Psi_0^\alpha$
then leads to the Taylor expansion
\begin{equation}
  \label{OmExp}
  \Omega\Psi_0^\alpha =\Big[\Ombar(E_0) +\partder{\Ombar}{\E}\Big|_{E_0}\;\Delta E^\alpha+
  \half\ppartder{\Ombar}{\E}\Big|_{E_0}\;(\Delta E^\alpha)^2
  +\frac{1}{3!}\cdots\Big]\Psi_0^\alpha=\Ombar(E^\alpha)\Psi_0^\alpha
\end{equation}
This implies that \it{the MSC term shifts the energy parameter of
the resolvent as well as that of the potential from the
unperturbed energy $E_0$ to the exact energy $E^\alpha$}. But
$\Ombar(E^\alpha)\,\Psi_0^\alpha$ with the energy parameter equal
to the full energy for the state $\Psi^\alpha$ is also identical
to the \it{Brillouin-Wigner expansion} \eq{BSBW},
\begin{equation}
  \label{OmBW}
  \Ombar(E^\alpha)\Psi_0^\alpha=\Big[1+\frac{Q}{E^\alpha-H_0}\,\V(E^\alpha)
  +\frac{Q}{E^\alpha-H_0}\,\V(E^\alpha)\frac{Q}{E^\alpha-H_0}\,\V(E^\alpha)
  +\cdots\Big]\Psi_0^\alpha\hsp
\end{equation}
which represents the full wave function, i.e.,
\begin{equation}
  \label{OmOmbar}
  \boxed{\Ombar(E^\alpha)\Psi_0^\alpha=\Om\Psi_0^\alpha}
\end{equation}
This implies that the relation \eq{Key} essentially represents
\it{the link between the Rayleigh-Schr\ödinger and the
Brillouin-Wigner expansions for an energy-dependent interaction}
and at the same time \it{the link between the MBPT approaches and
the Bethe-Salpeter equation} (indicated by the arrow in the
diagram of Fig. \ref{Fig:Intr}).

The BW expansion \eq{OmBW} can be expressed
\begin{equation}
  \label{BSA4}
  \Om\Psi_0^\alpha=\Psi_0^\alpha+\GQ(E^\alpha)\V(E^\alpha)
  \Om\Psi_0^\alpha
\end{equation}
or
\begin{equation}
  \label{BSA6}
  (E^\alpha-H_0)\,Q\Psi^\alpha=Q\calV(E^\alpha)\Psi^\alpha
\end{equation}

From the relation \eq{HHbar} it can be shown in analogy with the
relation \eq{Key}
\begin{equation}
  \label{HKey}
  H'\eff=\Hbar'\eff+\sum_{n=1}^\infty
 \npartder{n}{\Hbar'\eff}{\E}\;\big(H'\eff\big)^{n}
\end{equation}
With the definitions \eq{Hbar} and \eq{Ombar} this leads to
\begin{equation}
  \label{Hb}
  \Hbar'\eff=P\V(E_0)\,\Ombar(E_0) P
\end{equation}
and in analogy with the relation \eq{OmExp} to
\begin{equation}
  \label{Hef}
  H'\eff=P\V(E^\alpha)\,\Ombar(E^\alpha) P=P\V(E^\alpha)\,\Om P
\end{equation}
This leads together with Eq. \eq{BSA6} to the final equation
\begin{equation}
  \label{BSA}
  \boxed{(E^\alpha-H_0)\,\Psi^\alpha= \calV(E^\alpha)\,\Psi^\alpha}
\end{equation}
\textbf{This is the Bethe-Salpeter equation for energy-dependent
interactions in the Schr\ödinger-like form \eq{BS}}.

\it{We have now confirmed that the Schr\ödinger equation
\eq{GML1}, obtained directly from the generalized Gell-Mann--Low
relation in the extended Fock space with the perturbation
\eq{Pert}, corresponds in the projected Hilbert space with no
uncontracted photons to a Schr\ödinger-like equation with the
perturbation \eq{Vgen}. Both forms represent the complete
interaction between the particles and are exactly equivalent to
the original Bethe-Salpeter equation \eq{BS2}.}

The main difference between the original form of the BS equation
and the Schr\ödinger-like form derived here is primarily that the
latter has the time dependence reduced to a single time, which
makes the wave function in accord with standard quantum mechanics.
Furthermore, the Schr\ödinger-like form contains explicitly the
resolvent, while the remaining part of the Green's function
\eq{G0Op} is merged with the kernel $\kappa$ to form the potential
$\V$.

The Schr\ödinger-like equation \eq{BSA} we have derived is
equivalent to the equation derived from the BS equation by
Sucher~\cite[Eq. 1.47]{Su58} and rederived by Douglas and
Kroll~\cite[Eq. 3.26]{DK74} and Zhang~\cite[Eq. 15]{Zhang96}. In
these works the equation is essentially obtained by integrating
over the relative energy of the particles, thereby transforming
the equation to an "equal-times" equation. This equation is then
analyzed in terms of the Brillouin-Wigner perturbation theory. In
our presentation the corresponding equation is obtained by
starting from MBPT in the Rayleigh-Schr\ödinger formulation and
summing all relevant perturbations to all orders. The present
derivation therefore can serve as a link between the two
approaches.

In the next section we shall extend the treatment to the
quasi-degenerate case and derive the corresponding Bloch equation.

\subsection{Derivation of the Bethe-Salpeter-Bloch
equation. Quasi-degenerate model space.}
  \label{sec:BSB}

We have previously assumed that the model space is
\it{degenerate}, which for a two-electron system implies that the
effective interaction is \it{diagonal} within this space (assuming
the basis functions have definite symmetry). Then the relation
\eq{DeltaE} simplifies the treatment, and the formulas derived in
the previous section lead directly to the standard Bethe-Salpeter
equation \eq{BSA}. The treatment above, however, is more general
and can be extended to the case where the model space is
non-degenerate (quasi-degenerate). In the present section we shall
show how this can be performed.

The following relation can easily derived by induction
\begin{equation}
  \label{Der1}
  \partdern{n}{\Ombar}{\E}=\GQ\partdern{n}{(\V\Ombar)}{\E}
  -n\GQ\partdern{{(n-1)}}{\Ombar}{\E}
\end{equation}
To prove this we form the next-order derivative
  \[\partdern{{(n+1)}}{\Ombar}{\E}=-\GQ^2\partdern{n}{(\V\Ombar)}{\E}
  +\GQ\partdern{{(n+1)}}{(\V\Ombar)}{\E}
  +n\GQ^2\partdern{{(n-1)}}{\Ombar}{\E}-n\GQ\partdern{{n}}{\Ombar}{\E}\]
  (Since no singularities are involved here, ordinary rules of
  derivation can be used.)
Inserting the expression \eq{Der1} in the first term, yields
  \[\partdern{{(n+1)}}{\Ombar}{\E}=\GQ\partdern{{(n+1)}}{(\V\Ombar)}{\E}
  -(n+1)\GQ\partdern{{n}}{\Ombar}{\E}\]
In first order we have with $\Ombar=1-\GQ\V\Ombar$
  \[\partder{\Ombar}{\E}=\GQ\partder{(\V\Ombar)}{\E}-\GQ\Ombar\]
which completes the proof of the relation \eq{Der1}.

The formula above leads together with the expansion \eq{Key} to
\begin{equation}
  \label{Key1}
  Q\Om P=Q\Ombar P+\GQ\sum_{n=1}^\infty
 \npartder{n}{(\V\Ombar)}{\E}\;(H'\eff)^n-\GQ\sum_{n=1}^\infty
 \npartder{{(n-1)}}{\Ombar}{\E}\;(H'\eff)^n
\end{equation}
The first term on the rhs can also be expressed $\GQ\V\Ombar P$,
and the last term is simply $-\GQ\Om H'\eff$, which yields
\begin{equation}
  \label{Key2}
  Q\Om P=\GQ\V\Ombar P+\GQ\sum_{n=1}^\infty
 \npartder{n}{(\V\Ombar)}{\E}\;(H'\eff)^n-\GQ\Om H'\eff
\end{equation}

We can consider $\V\Ombar$ as a single energy-dependent operator,
and if that operates on a particular model state of a degenerate
model space of energy $E_0$, the first two terms of the bracket
above represents the Taylor expansion
\begin{equation}
  \label{Taylor}
  \V(E_0)\,\Ombar(E_0)+\sum_{n=1}^\infty
 \npartder{n}{(\V\Ombar)}{\E}\Big|_{E_0}\;(\Delta E^\alpha)^n
 =\V(E^\alpha)\,\Ombar(E^\alpha).\hsp
\end{equation}
Thus, the expansion has the effect of transforming the energy
parameter of the product $\V\Ombar$ from $E_0$ to the full energy
$E^\alpha$,
\begin{equation}
  \label{ModE}
  \V(E_0)\Ombar(E_0)\Psi_0^\alpha\rarr\V(E^\alpha)\Ombar(E^\alpha)\Psi_0^\alpha
\end{equation}
in analogy with the expansion \eq{OmBW}. Using the relation
\eq{OmOmbar}, the equation \eq{Key2} above then becomes
\begin{equation}
  \label{BSB1}
  Q\Om \Psi_0^\alpha=\GQ\big[\V(E^\alpha)\Om-\Om
  H\eff'\big]\Psi_0^\alpha
\end{equation}
or
\begin{equation}
  \label{BSB2}
  (E_0-H_0)\Om \Psi_0^\alpha=Q\big[\V(E^\alpha)\Om-\Om
  H\eff'\big]\Psi_0^\alpha
\end{equation}
which is consistent with the Bethe-Salpeter equation \eq{BSA}.

If the model space is \it{non-degenerate} (quasi-degenerate), then
the relation \eq{DeltaE} is no longer valid, and the expansion
\eq{Key2} can not be expressed by means of a single energy
parameter as in the Taylor expansion \eq{Taylor}. Instead, the
potential will depend on the \it{full matrix} of the effective
Hamiltonian. We then replace the energy parameter in \eq{OmbarE0}
by the model Hamiltonian $H_0$,
  \[\Ombar P=P+\GQ(H_0)\V(H_0)+\cdots=\Ombar(H_0)P\]
By this notation we understand---in accordance with the rule
\eq{OmbarE}---
\begin{equation}
  A(H_0)B\Phi=A(E_0)B\Phi
\end{equation}
when $\Phi$ is an eigenfunction of $H_0$ with the eigenvalue $E_0$
and $B$ is an arbitrary operator combination. Together with the
linearity condition,
\begin{equation}
  A(H_0)B(a\Phi+b\Phi'\big)=aA(E_0)B\Phi+bA(E'_0)B\Phi'
\end{equation}
where $\Phi'$ is another eigenfunction of $H_0$ with the
eigenvalue $E'_0$, this defines the notation fully.

The expansion \eq{Key2} can now be regarded, in analogy with the
energy modification \eq{Taylor}, as modifying the parameter $H_0$
to the full effective Hamiltonian $H\eff=H_0+H'\eff$
\begin{equation}
  \label{Taylor2}
  \V(H_0)\,\Ombar(H_0)+\sum_{n=1}^\infty
 \npartder{n}{(\V\Ombar)}{\E}\;(H'\eff)^n
 =\V(H\eff)\,\Ombar(H\eff)\hsp
\end{equation}
i.e.,
\begin{equation}
  \label{ModH}
  \V(H_0)\Ombar(H_0)P\rarr\V(H\eff)\Ombar(H\eff)P
\end{equation}
and Eq. \eq{BSB1} becomes
\begin{equation}
  \label{VOm2}
  Q\Om P=\GQ(H_0)\V(H\eff)\Ombar(H\eff)P-\GQ(H_0)\Om H'\eff
\end{equation}
The notation here is defined by the relation
\begin{equation}
  A(H\eff)B\Psi_0^\alpha=A(E^\alpha)B\Psi_0^\alpha
\end{equation}
where $\Psi_0^\alpha$ is a model function (eigenfunction of
$H\eff$ with the eigenvalue $E^\alpha$, see Eq. \ref{SecEq}),
which together with the linearity condition defines the operator
when acting on any model space.

Similarly, the expansion \eq{Key} yields
\begin{equation}
  \label{Keyref}
  \Om P=\Ombar(H_0)P+\sum_{n=1}^\infty
  \npartder{n}{\Ombar}{\E}\;\big(H'\eff\big)^{n}=\Ombar(H\eff)P
\end{equation}
and we can now express the equation \eq{VOm2} as
\begin{equation}
  Q\Om P=\GQ(H_0)\V(H\eff)\,\Om P-\GQ(H_0)\Om\,H\eff'
\end{equation}
or
\begin{equation}
  \GQ(H_0)]^{-1}Q\Om P=Q\big[\V(H\eff)\,\Om-Q\Om\,H\eff'\big]P
\end{equation}
Operating on a model-state function $\Phi(E_0^\alpha)$ of energy
$E_0^\alpha$, we have according to the definitions above
  \[[\GQ(H_0)]^{-1}\Phi(E_0^\alpha)=(E_0^\alpha-H_0)\,\Om\Phi(E_0^\alpha)\]
Therefore, \it{the inverse of the resolvent can be expressed as a
commutator}
\begin{equation}
  \label{comm}
  [\GQ(H_0)]^{-1}A\equiv[A,H_0]
\end{equation}
where $A$ is an arbitrary operator. This leads to the commutator
relation
\begin{equation}
  \label{BScomm}
  [\Om,H_0] P=Q\V(H\eff)\,\Om P-Q\Om\,H\eff'
\end{equation}
The relation \eq{Hef} can be generalized to
\begin{equation}
  \label{HeffGen}
  H\eff'=P\V(H\eff)\,\Om P
\end{equation}
and with the IN relation \eq{OmIN} $P\Om P=P$ we arrive at \it{the
BS equation in commutator form}
\begin{equation}
  \label{BSBloch}
  \boxed{\big[\Om,H_0\big]P=\V(H\eff)\,\Om P-\Om\,H\eff'}
\end{equation}

\it{The equation above---which we shall refer to as the
\textbf{Bethe-Salpeter-Bloch equation}---is the main result of the
present work. It has the same relation to the standard BS equation
as has the standard Bloch equation \eq{Bloch} to the ordinary
Schr\ödinger equation. It can be used to generate the BS equation
perturbatively, essentially as the ordinary Bloch eqution is used
in standard MBPT. The commutator form makes it possible to apply
the equation to an extended model space, which essentially
eliminates the quasi-degeneracy problem that might appear in
applying the standard equation directly on a single state.}

\subsection{Expansion of the Bethe-Salpeter-Bloch equation}

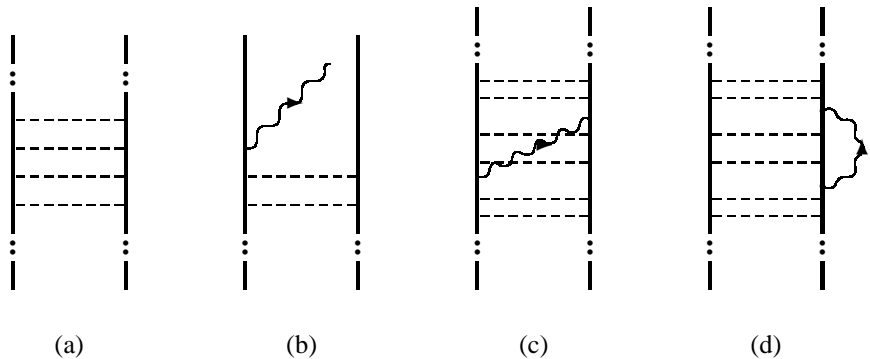
\begin{figure}[hbt]
\begin{center}
\begin{picture}(4,6)(-1,-1)
 \put(0,3.67){\circle*{0.1}}\put(2,3.67){\circle*{0.1}}
 \put(0,3.83){\circle*{0.1}}\put(2,3.83){\circle*{0.1}}
 \put(0,4){\lline{0.5}{0.5}{}{}}
 \put(2,4){\lline{0.5}{0.5}{}{}}
 \put(0,3){\elstat{}{}{}}
 \put(0,2.5){\elstat{}{}{}}
 \put(0,2){\elstat{}{}{}}
 \put(0,1.5){\elstat{}{}{}}
 \put(0,1){\lline{2.5}{0.5}{}{}}
 \put(2,1){\lline{2.5}{0.5}{}{}}
 \put(0,0.67){\circle*{0.1}}\put(2,0.67){\circle*{0.1}}
 \put(0,0.83){\circle*{0.1}}\put(2,0.83){\circle*{0.1}}
 \put(0,0){\lline{0.5}{0.5}{}{}}
 \put(2,0){\lline{0.5}{0.5}{}{}}
 \put(1,-1){\makebox(0,0){(a)}}
 \end{picture}
\begin{picture}(4,6)(-1,-1)
 \put(0,4){\lline{0.5}{0.5}{}{}}
 \put(2,4){\lline{0.5}{0.5}{}{}}
 \put(0,2.5){\photonNe{}{}{}}
 \put(0,2){\elstat{}{}{}}
 \put(0,1.5){\elstat{}{}{}}
 \put(0,1){\lline{3.5}{0.5}{}{}}
 \put(2,1){\lline{3.5}{0.5}{}{}}
 \put(0,0.67){\circle*{0.1}}\put(2,0.67){\circle*{0.1}}
 \put(0,0.83){\circle*{0.1}}\put(2,0.83){\circle*{0.1}}
 \put(0,0){\lline{0.5}{0.5}{}{}}
 \put(2,0){\lline{0.5}{0.5}{}{}}
 \put(1,-1){\makebox(0,0){(b)}}
 \end{picture}
 \begin{picture}(4,6)(-1,-1)
 \put(0,4.17){\circle*{0.1}}\put(2,4.17){\circle*{0.1}}
 \put(0,4.33){\circle*{0.1}}\put(2,4.33){\circle*{0.1}}
 \put(0,4.5){\lline{0.5}{0.5}{}{}}
 \put(2,4.5){\lline{0.5}{0.5}{}{}}
 \put(0,3.4){\elstat{}{}{}}
 \put(0,3.7){\elstat{}{}{}}
 \put(0,2){\photonENE{}{}{}}
 \put(0,2.75){\elstat{}{}{}}
 \put(0,2.25){\elstat{}{}{}}
 \put(0,1.6){\elstat{}{}{}}
 \put(0,1.3){\elstat{}{}{}}
 \put(0,1){\lline{3}{0.5}{}{}}
 \put(2,1){\lline{3}{0.5}{}{}}
 \put(0,0.67){\circle*{0.1}}\put(2,0.67){\circle*{0.1}}
 \put(0,0.83){\circle*{0.1}}\put(2,0.83){\circle*{0.1}}
 \put(0,0){\lline{0.5}{0.5}{}{}}
 \put(2,0){\lline{0.5}{0.5}{}{}}
 \put(1,-1){\makebox(0,0){(c)}}
 \end{picture}
  \begin{picture}(4,6)(-1,-1)
 \put(0,4.17){\circle*{0.1}}\put(2,4.17){\circle*{0.1}}
 \put(0,4.33){\circle*{0.1}}\put(2,4.33){\circle*{0.1}}
 \put(0,4.5){\lline{0.5}{0.5}{}{}}
 \put(2,4.5){\lline{0.5}{0.5}{}{}}
 \put(0,3.4){\elstat{}{}{}}
 \put(0,3.7){\elstat{}{}{}}
 {\setlength{\unitlength}{0.5cm}
 \put(3,3.75){\ElSE{}{}{}}}
 \put(0,2.75){\elstat{}{}{}}
 \put(0,2.25){\elstat{}{}{}}
 \put(0,1.6){\elstat{}{}{}}
 \put(0,1.3){\elstat{}{}{}}
 \put(0,1){\lline{3}{0.5}{}{}}
 \put(2,1){\lline{3}{0.5}{}{}}
 \put(0,0.67){\circle*{0.1}}\put(2,0.67){\circle*{0.1}}
 \put(0,0.83){\circle*{0.1}}\put(2,0.83){\circle*{0.1}}
 \put(0,0){\lline{0.5}{0.5}{}{}}
 \put(2,0){\lline{0.5}{0.5}{}{}}
 \put(1,-1){\makebox(0,0){(d)}}
 \end{picture}
    \renewcommand{\normalsize}{\footnotesize}
    \caption{Illustration of the perturbative-nonperturbative
    procedure for solving the Bethe-Salpeter equation, described in the text.}
    \renewcommand{\normalsize}{\standard}
    \label{Fig:Pair}
\end{center}
\end{figure}

The original Bethe-Salpeter equation contains the exact energy and
is therefore normally treated by means a Brillouin-Wigner
perturbation expansion~\cite{DK74,Zhang96a}, which requires a
self-consistent treatment. The BS equation in the Bloch-equation
form can be used to generate a perturbation expansion of
Rayleigh-Schr\ödinger type that does not require any
self-consistence procedure. We have at our laboratory developed a
procedure of solving the Bethe-Salpeter-Bloch equation that is a
combination of perturbative and non-perturbative techniques, which
we shall here briefly indicate. A detailed description of the
procedure together with numerical results will appear
shortly~\cite{HSL05}.

Our procedure is based upon the iterative solution of pair
equations~\cite{Ma79,ELi88,SO89,SO89a,LM86}. This represents the
"ladder" approximation of the BS equation, as indicated in Fig.
\ref{Fig:Pair} (a). The pair function is combined with the
emission of a single (uncontracted) photon (Fig. \ref{Fig:Pair}
b). This represents a function in the extended Fock space,
discussed in section \ref{sec:Wop}. This function can be iterated
further, before the photon is annihilated, which can yield
instantaneous (Coulomb and Breit) interactions, crossing the
photon. These iterations can be continued after the annihilation,
as indicated in Fig. \ref{Fig:Pair} (c). By annihilating the
photon on the same electron line, leads to self-energy and vertex
corrections (Fig. \ref{Fig:Pair} d). At present time it is
possible to treat only one covariant photon in this way, but the
dominating part of the multi-photon exchange will be included by
the crossings with the instantaneous interactions (c). This will
correspond to all effects treated by Zhang~\cite{Zhang96a} in his
analysis of the helium fine structure up to order $m\alpha^7$,
except for the non-separable part of two-photon exchange (Fig.
\ref{Fig:NonSep}). With our approach this part has at present to
be included analytically.

\section{Summary and conclusions}
Standard many-body perturbation theory (MBPT) is conveniently
based upon the Bloch equation, which is the generating equation
for Rayleigh-Schr\ödinger perturbation expansion. The Bloch
equation can also be used to generate various other perturbative
schemes, such as the linked-diagram expansion, and it also leads
to non-perturbative schemes, such as the Coupled-Cluster Approach.
In the commutator form \eq{Bloch} the Bloch equation leads to
schemes that can handle the quasi-degenerate problem in an
efficient way by means of an "extended" model space.

In this paper we have reviewed the connection between relativistic
MBPT and quantum-electrodynamics (QED) for a two-electron system
by means of the recently introduced covariant-evolution-operator
method~\cite{LSA04}. The exchange of a single covariant photon is
treated to all orders, and this is shown to lead to an equation of
the Bethe-Salpeter (BS) type. Extending the treatment to all
non-separable interactions (including radiative corrections) leads
to the full BS equation. This establishes a link between the
perturbative and non-perturbative schemes, based upon
Rayleigh-Schr\ödinger perturbation theory and schemes based upon
the BS equation, which are normally treated by means of
Brillouin-Wigner perturbation theory.

In addition, a Bloch equation in commutator form that is
compatible with the BS equation is derived. This equation has the
same relation to the Bethe-Salpeter equation as has the standard
Bloch equation to the ordinary Schr\ödinger equation and
represents a series of BS equations, associated with a model space
that need not be degenerate. This can be used to generate a
perturbative expansion, corresponding to the BS equation for an
extended model space. In principle, this will make it possible to
treat the quasi-degeneracy problem also within the BS formalism.
Such a scheme is presently being tested at our laboratory.

\section*{Acknowledgements}
The author wish to express his gratitude to Sten Salomonson,
Ann-Marie Pendrill, Bj\örn \Ås\'en and Daniel Hedendahl for the
collaboration on which the work presented here is largely based.
Stimulating discussions with Gordon Drake has been particular
valuable for this work and gratefully acknowledged.

\newpage\appendix
\begin{center}\textbf{APPENDIX}
\end{center}
\subsection{Zeroth-order Green's function}
\label{sec:GF}
                                   \renewcommand{\calE}{\mathcal{E}}
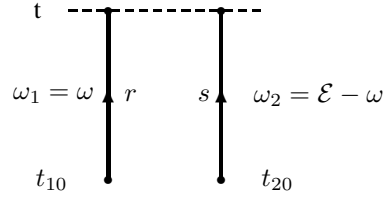
\begin{figure}[t]
\begin{center}
\begin{picture}(5,3.5)(-1.5,0)
 \put(-0.75,3){\multiput(0.05,0)(0.25,0){14}{\line(1,0){0.15}}}
 \put(-1.25,3){\makebox(0,0){t}}
 \put(0,3){\circle*{0.15}}
 \put(2,3){\circle*{0.15}}
  \put(0,0){\Elline{3}{1.5}{\omega_1=\omega\Hsp}{r}}
 \put(2,0){\Elline{3}{1.5}{s}{\Hsp\Hsp\omega_2= \calE-\omega}}
  \put(0,0){\circle*{0.15}}
 \put(2,0){\circle*{0.15}}
 \put(-1,0){\makebox(0,0){$t_{10}$}}
 \put(3,0){\makebox(0,0){$t_{20}$}}
  \end{picture}
 \vsp\caption{Graphical representation of the zeroth-order Green's function \eq{G0C}.} \label{Fig:G0}
 \end{center}
 \end{figure}

The zeroth-order Green's function \eq{GF0} in Fig. \ref{Fig:G0} is
in coordinate representation
\begin{equation}
  \label{G0C}
  G_0(x_1x_2;x_{10}x_{20})=\SF(x_1;x_{10})\,\SF(x_2;x_{20})
\end{equation}
where $\SF$ is the electron propagator
\begin{equation}
  \label{S}
  S(x,x_0)=\intd{\omega}\,S(\omega)\,e^{-\im\omega(t-t_0)}
\end{equation}
with the fourier transform
\begin{equation}
  \label{Sft}
  \bra{\bx}S(\omega)\ket{\bx_0}=\frac{\langle{\bx}\ket{r}\bra{r}{\bx_0}\rangle}
  {\omega-\eps_r+\im\eta_r}=\bra{\bx}\hat{S}(\omega)\ket{\bx_0}
\end{equation}
and the corresponding operator form
\begin{equation}
  \label{Sop}
  \hat{S}(\omega)=  \frac{\Lambda_+}{\omega-h+\im\eta}+\frac{\Lambda_-}{\omega-h-\im\eta}
\end{equation}
Here, $h$ is the single-electron Dirac Hamiltonian in the field of
the nucleus and $\Lambda_\pm$ are projection operators for
positive and negative-energy single-particle states.

We consider the equal-times Green's function with $t_1=t_2=t$,
which gives
\begin{eqnarray}
  \label{G0E}
  \Hsp&& G_0(t,\bx_1,\bx_2;x_{10},x_{20})=
  \intd{\epsi}\,\me^{-\im\epsi t}\,
  \frac{\langle\bx_1\bx_2\ket{rs}\bra{rs}\bx_{10}\bx_{20}\rangle}
  {\epsi-\eps_r-\eps_s}\nn
  \Hsp&\times&  \intd{\omega}\;\Big[\frac{1}{\omega-\eps_r+\im\eta_r}+
  \frac{1}{\epsi-\omega-\eps_s+\im\eta_s}\Big]\,\me^{\im\omega
  t_{10}}\,\me^{\im(\epsi-\omega)t_{20}}\Hsp
\end{eqnarray}
with $x=(t,\bx)$, $\omega_1=\omega$ and $\epsi=\omega_1+\omega_2$.
The fourier transform with respect to $t$ is then
\begin{eqnarray}
  \label{G0ft}
  &&G_0(\epsi,\bx_1,\bx_2;x_{10},x_{20})=\nn
  &&\frac{\langle\bx_1\bx_2\ket{rs}\bra{rs}\bx_{10}\bx_{20}\rangle}
  {\epsi-\eps_r-\eps_s}  \intd{\omega}\;\Big[\frac{1}{\omega-\eps_r+\im\eta_r}+
  \frac{1}{\epsi-\omega-\eps_s+\im\eta_s}\Big]\;
  \me^{\im\omega t_{10}}\,\me^{\im(\epsi-\omega)t_{20}}\Hsp
\end{eqnarray}
or in operator form
\begin{equation}
  \label{G0Op}
  G_0(\epsi)=\G(\epsi)\intd{\omega} \; g_0(\epsi,\omega)\,\me^{\im\omega t_{10}}\,
  \me^{\im({\epsi}-\omega)t_{20}}
\end{equation}
where $\G(\E)$ is the resolvent \eq{Gamma1}
\begin{equation}
  \label{Gamma}
  \G(\E)=\frac{1}{\E-H_0}
\end{equation}
$H_0=h_1+h_2$ is the zeroth-order Hamiltonian \eq{Pert} and
\begin{eqnarray}
  \label{g0}
  \Hsp g_0(\epsi,\omega)&=&\Lambda_+\Big[\frac{1}{\omega-h_1+\im\eta}
 +\frac{1}{\epsi-\omega-h_2+\im\eta}\Big]\nn
  \Hsp&+&\Lambda_-\Big[\frac{1}{\omega-h_1-\im\eta}
  +\frac{1}{\epsi-\omega-h_2-\im\eta}\Big]
\end{eqnarray}
 The inverse transformation is
\begin{equation}
  \label{G0ftint}
  \boxed{G_0(t,\bx_1,\bx_2;x_{10},x_{20})=
  \intd{\epsi}\;\me^{-\im\epsi
  t}\,G_0(\epsi,\bx_1,\bx_2;x_{10},x_{20})}
\end{equation}
and specifically,
\begin{equation}
  \label{G0ft0}
  \boxed{G_0(t=0,\bx_1,\bx_2;x_{10},x_{20})=
  \intd{\epsi}\;G_0(\epsi,\bx_1,\bx_2;x_{10},x_{20})}
\end{equation}

\subsection{Single-photon exchange}
\label{App:SingPhot} (See Ref.~\cite[Eq. 312]{LSA04}.) We consider
now the covariant evolution operator \eq{CovEv} for the exchange
of a single covariant photon, represented by the diagram in Fig.
\ref{Fig:SingPhot} (left)
\begin{eqnarray}
  \label{SP}
  U\ett(t',t_0)&=&-\half\dint \dif^3\bx'_1\,\dif^3\bx'_2\,\hpsi\dagg(x'_1)\hpsi\dagg(x'_2)
  \dint \dif^4 x_1\,\dif^4 x_2\,\im\SF(x'_1,x_1)\,\im\SF(x'_2,x_2)\,\im  I(x_2,x_1)\nn
  &\times&\dint\dif^3\bx_{10}\,\dif^3\bx_{20}\,\im\SF(x_1,x_{10})\,\im\SF(x_2,x_{20})
  \,\hpsi(x_{20})\hpsi(x_{10})
\end{eqnarray}
leaving out the damping factors. More compactly, we express this
as
\begin{eqnarray}
  \label{SPa}
   \Hsp U\ett(t',t_0)&=&-\half\hpsi\dagg(x'_1)\hpsi\dagg(x'_2)
  \,  G_0(x'_1,x'_2;x_1x_2)\,\im  I(x_2,x_1)\nn
  \Hsp&\times&G_0(x_1,x_2;x_{10}x_{20})\;\hpsi(x_{20})\hpsi(x_{10})
\end{eqnarray}
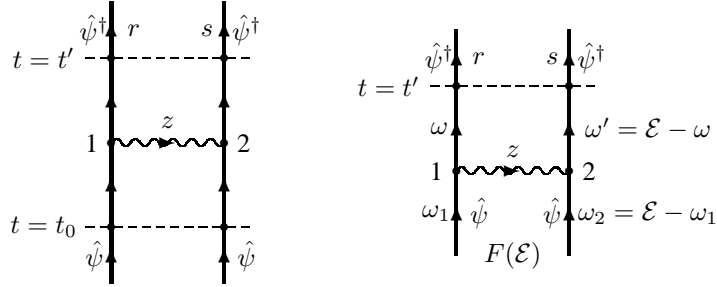
\begin{figure}[hbt]
\begin{center}
\begin{picture}(6,6)(-1,-1.5)
\put(-0.5,3){\dash{12}} \put(-1.2,3){\makebox(0,0){$t=t'$}}
\put(0,3){\Elline{1}{0.5}{\hpsi\dagg}{r}}
\put(2,3){\Elline{1}{0.5}{s}{\hpsi\dagg}}
\put(0,0){\Elline{1.5}{0.75}{}{}}
\put(2,0){\Elline{1.5}{0.75}{}{}} \put(0,1.5){\photon{z}{1}{2}}
 \put(0,1.5){\Elline{1.5}{0.75}{}{}}
 \put(2,1.5){\Elline{1.5}{0.75}{}{}}
\put(0,3){\circle*{0.15}} \put(2,3){\circle*{0.15}}
 \put(-0.5,0){\dash{12}} \put(-1.2,0){\makebox(0,0){$t=t_0$}}
 \put(0,0){\circle*{0.15}} \put(2,0){\circle*{0.15}}
\put(0,-1){\Elline{1}{0.5}{\hpsi}{}}
\put(2,-1){\Elline{1}{0.5}{}{\hpsi}}
\end{picture}
\begin{picture}(6,6)(-1,-1)
\put(-0.5,3){\dash{12}} \put(-1.2,3){\makebox(0,0){$t=t'$}}
\put(0,3){\Elline{1}{0.5}{\hpsi\dagg}{r}}
\put(2,3){\Elline{1}{0.5}{s}{\hpsi\dagg}}
\put(0,0){\Elline{1.5}{0.75}{\omega_1\:}{\hpsi}}
\put(2,0){\Elline{1.5}{0.75}{\hpsi}{\hsp\Hsp
\omega_2=\E-\omega_1}} \put(0,1.5){\photon{z}{1}{2}}
 \put(0,1.5){\Elline{1.5}{0.75}{\omega}{}}
 \put(2,1.5){\Elline{1.5}{0.75}{}{\hsp\Hsp \omega'=\E-\omega}}
\put(0,3){\circle*{0.15}} \put(2,3){\circle*{0.15}}
 \put(1,0){\makebox(0,0){$F(\calE)$}}
\end{picture}
    \renewcommand{\normalsize}{\footnotesize}
    \caption{Graphical representation of the covariant-evolution
      operator for single-photon exchange in the form \eq{SP} (left) and in the
      form \eq{SPb} with $t_0\rarr-\infty$ (right).}
    \renewcommand{\normalsize}{\standard}
    \label{Fig:SingPhot}
\end{center}
\end{figure}
with integrations over all variables that do not appear on the
left-hand side. Here, $I(x_2,x_1)$ represents the single-photon
exchange
\begin{equation}
  \label{I}
  I(x_2,x_1)= e\alpha_1^{\mu} \DF(x_2-x_1)\, e\alpha_2^{\nu}=
  \intd{z}\;\me^{-\im z(t_2-t_1)}\,\im I(z,\bx_2,\bx_1)
\end{equation}
where $\DF(x_2-x_1)$ is the \it{Feynman photon propagator}.

If we operate to the right on a \it{positive-energy state}, we can
use the relations \eq{Rel} to simplify the expression.
Furthermore, since in that case $t_0\leq t_1, t_2$ and since $t_1,
t_2$ run from $-\infty$ to $+\infty$, we must have $t_0=-\infty$,
yielding
\begin{equation}
  \label{SPb}
  U\ett(t',-\infty)=\half\hpsi\dagg(x'_1)\hpsi\dagg(x'_2)
  \,  G_0(x'_1,x'_2;x_1x_2)\,\im
  I(x_2,x_1)\;\hpsi(x_{2})\hpsi(x_{1})
\end{equation}

The electron-field operator is in the interaction
picture~\cite{FW71}
\begin{equation}
  \label{Elfield}
  \hpsi(x)=\hpsi(t,\bx)=c_j\phi_j(\bx)\,e^{-\im\eps_j t}
\end{equation}
with the fourier transform
\begin{equation}
  \label{ElFT}
  \hpsi(\omega,\bx)=\int\dif t\,e^{\im\omega t}\,\hpsi(x)
  =c_j\phi_j(\bx)\,2\pi\delta(\omega-\eps_j)
\end{equation}
(as usual, summed over repeated indices) and the inverse transform
\begin{equation}
  \label{ElFTinv}
  \hpsi(x)=\intd{\omega}\,e^{-\im\omega t}\,\hpsi(\omega,\bx)
\end{equation}
An arbitrary function of $x_1=(t_1,\bx_1)$ can be fourier expanded
as
\begin{equation}
  \label{F}
  F(x_1)=\intd{\omega_1}\;e^{-\im t_1\omega_1}
  \,F(\omega_1,\bx_1)
\end{equation}
Operating on a (time-independent) fourier component of that
function with the electron-field operator \eq{ElFTinv}, yields
\begin{equation}
  \label{PsiF}
  \hpsi(x)\,F(\omega_1,\bx_1)=\intd{\omega}\,e^{-\im\omega t}\,\hpsi(\omega,\bx)
  \,F(\omega_1,\bx_1)=e^{-\im\omega_1 t} \,F(\omega_1,\bx_1)
\end{equation}
With the adiabatic damping the time-independent component
corresponds to $t_1=-\infty$, which implies that the field
operator propagates the function $F(\omega_1,\bx_1)$ from the time
$-\infty$ to $t$. Similarly, the product of two electron-field
operators $\hpsi(x_1)\hpsi(x_2)$ operating on a time-independent
two-electron function, propagates the individual electrons from
the time $t=-\infty$ to $t=t_1$ and $t=t_2$, respectively,
\it{without} any electron-electron interaction. Thus,
\begin{equation}
  \label{FT}
  \hpsi(x_1)\hpsi(x_2)\,F(\omega_1,\omega_2)=
  e^{-\im t_1\omega_1}\,e^{-\im t_2\omega_2}
  \,F(\omega_1,\omega_2)
\end{equation}

We now use the form \eq{G0ftint} of the Green's function
\begin{equation}
  \label{GFSP}
  G_0(x_1',x_2';x_1,x_2)=\intd{\epsi}\;e^{-\im\epsi t'}\,G_0(\epsi)
  =\intd{\epsi}\;e^{-\im\epsi t'}\,\G(\epsi)\intd{\omega}\,g_0(\epsi,\omega)\,
  e^{\im\omega t_1}\,e^{\im(\epsi-\omega)t_2}
\end{equation}
to operate with the evolution operator \eq{SPb} on the fourier
component $F(\omega_1,\omega_2)$ (see also Fig.
\ref{Fig:SingPhot}, right), which yields
\begin{eqnarray}
  &&U\ett(t',-\infty)\,F(\omega_1,\omega_2)=
  \half c_j\dagg\phi_j\dagg(\bx_1)\,e^{\im t'\eps_j}
  \,c_k\dagg\phi_k\dagg(\bx_2)\,e^{\im t'\eps_j}
\intd{\epsi}\,\G(\epsi)\,e^{-\im t'\epsi}\nn
  &\times& \intd{\omega}\,g_0(\epsi,\omega)\,  \intd{z}\;\im I(z)
  e^{-\im t_1(\omega_1-z-\omega)}\,e^{-\im
t_2(\omega_2+z-\epsi+\omega)}\,F(\omega_1,\omega_2)
\end{eqnarray}
and after time integrations
\begin{eqnarray}
  \label{SP2}
  &&U\ett(t',-\infty)\,F(\omega_1,\omega_2)=
  \ket{rs}\bra{rs}\intd{\epsi}\,e^{-\im t'(\epsi-\eps_r-\eps_s)}
 \G(\epsi)\intd{\omega}\,g_0(\epsi,\omega)\,\intd{z}\;\im I(z)  \nn
  &\times&F(\omega_1,\omega_2)
  \;2\pi\delta(\omega_1-z-\omega)\, 2\pi\delta(\omega_1+\omega_2-\epsi)
\end{eqnarray}
Here, $\ket{ij}$ represents a straight (non-symmetrized) product
of time-independent single-electron functions (which eliminates
the factor of $\halfS$).

If we operate on a particular energy component
\begin{equation}
  \label{FE}
  F(\E)=\intd{\omega_1}\;F(\omega_1,\E-\omega_1)
\end{equation}
the result becomes in operator form
\begin{eqnarray}
  \label{SP3}
  \Hsp U\ett(t',-\infty)\,F(\calE)&=&
  e^{-\im t'(\calE-H_0)}\G(\E)\intd{\omega}\,g_0(\calE,\omega)\nn
  \Hsp&\times&\intd{\omega_1}\;\im
  I(\omega_1-\omega)\,F(\omega_1,\calE-\omega_1)
\end{eqnarray}
We can also express this result as
\begin{equation}
  \label{SPc}
  U\ett(t,-\infty)\,F(\E)=e^{-\im  t(\E-H_0)}\,\G(\E)\,V(\E)\,F(\E)
\end{equation}
where
\begin{equation}
  \label{VE}
  V(\calE)=\intd{\omega}\;g_0(\E,\omega)
  \intd{\omega_1}\;\im I(\omega_1-\omega)
\end{equation}
Here, $\G(\E)$ is the resolvent \eq{Gamma} and $g_0(\calE,\omega)$
is the operator \eq{g0}. The corresponding effective interaction
is obtained from the relation \eq{EffInt} by taking the time
derivative at $t=0$, which eliminates the resolvent,
\begin{equation}
  H\eff\ett(\E)=PV(\E)\,P
\end{equation}

With the explicit form of the interaction \eq{I} the matrix
elements of the potential for the exchange of a single covariant
photon becomes~\cite[App. A]{LSA04}
\begin{equation}
  \label{VExpl}
  \bra{rs}V(\calE)\ket{tu}=\Bigbra{rs}\int f(k)\dif k\Big[\frac{1}
  {\E-\eps_r-\eps_u-(k-\im\gamma)_r}
  +\frac{1}{\E-\eps_s-\eps_t-(k-\im\gamma)_s}\Big]\Bigket{tu}
\end{equation}
where the $A_r=A\;\sgn(\eps_r)$. The function $f(k)$ is in the
Feynman gauge given by~\cite[Eq. 77]{LSA04}
\begin{eqnarray}
  \label{fk}
  f(k)&=&-\frac{e^2}{4\pi^2}\,(1-\bs{\alpha_1\cdot\alpha_2})\,
  \frac{\sin(k\rr)}{\rr}\nn
  &=&-\frac{e^2}{4\pi^2}\,(1-\bs{\alpha_1\cdot\alpha_2})\,
  \sum_{l=0}^\infty (2l+1)j_l(kr_1)j_l(kr_2)\,\bs{C}^{(k)}(1)\bsdot
  \bs{C}^{(k)}(2)
\end{eqnarray}
where $j_l$ are spherical Bessel functions and $\bs{C}^{(k)}$
spherical tensors, closely related to the spherical harmonics.

Summarizing, the contribution to the wave operator \eq{wf} from
single-photon exchange, when operating to the right on a function
of the type \eq{FE}, becomes
\begin{equation}
  \label{Om1}
  \boxed{\Omega\ett(\calE)=Q\Util\ett(0,-\infty)= \GQ(\calE) V(\calE)}
\end{equation}
where
\begin{equation}
  \label{GammaQ}
  \GQ(\E)=Q\G(\E)=\frac{Q}{\E-H_0}
\end{equation}
 and the corresponding contribution to the effective Hamiltonian
\begin{equation}
  \label{Heffett}
  \boxed{H\eff\ett(\E)=PV(\E)\,P}
\end{equation}
                                          \newcommand{\deltaE}{\delta \calE}

\subsection{Separable two-photon exchange}
  \label{App:SepPh}
  (See Ref.~\cite[Sect.5.2.1 and App.A.2]{LSA04})\\
Next we consider the separable two-photon exchange for which there
is an intermediate time $t=t"$ with no free or uncontracted
photons, as illustrated in Fig. \ref{Fig:SepPh} (left). The
evolution operator \eq{EvolOp} can in this case be expressed
\begin{equation}
  \label{U2a}
  U\tva(t,-\infty)P=U\ett(t,t")\,U\ett(t",-\infty)P
\end{equation}
Here, the intermediate states run over \it{all} states --- in the
$Q$ as well as the $P$ space --- and when the intermediate state
lies in the model space ($P$), \it{(quasi)singularities} may
occur. These singularities are removed in the \it{reduced
evolution operator} \eq{Ured} by including \it{counterterms}
\eq{C}
\begin{equation}
  \label{U+C}
  \Util(t)P=U(t)P+C(t)P
\end{equation}
We also recall the definition of the \it{model-space contribution}
(MSC) \eq{Mdef}
\begin{equation}
  \label{M}
  \Util(0)P=\Ubar(0)P+MP
\end{equation}
where $\Ubar$ is the evolution operator \eq{Ubar} with no
intermediate model-space state, in this case
\begin{equation}
  \label{Ubar2}
  \Ubar\tva P=\G V\GQ VP
\end{equation}
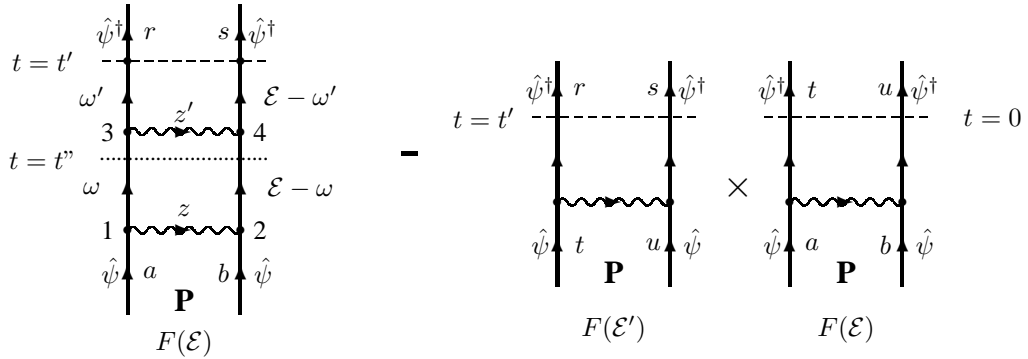
\begin{figure}[htb]
\begin{center}
\begin{picture}(7,6)(-1.5,-0.5)
\put(-0.5,4.5){\dash{12}}
\multiput(-0.5,2.75)(0.1,0.){30}{\small.}
\put(-1.5,4.5){\makebox(0,0){$t=t'$}}
\put(-1.5,2.75){\makebox(0,0){$t=t"$}}
\put(0,4.5){\Elline{1}{0.5}{\hpsi\dagg}{r}}
\put(2,4.5){\Elline{1}{0.5}{s}{\hpsi\dagg}}
\put(0,3.25){\Elline{1.25}{0.625}{\omega'\hsp}{}}
\put(2,3.25){\Elline{1.25}{0.625}{}{\Hsp\calE-\omega'}}
\put(0,0){\Elline{1.5}{0.75}{\hpsi}{a}}
\put(2,0){\Elline{1.5}{0.75}{b}{\hpsi}}
\put(0,1.5){\photon{z}{1}{2}} \put(0,3.25){\photon{z'}{3}{4}}
\put(0,1.5){\Elline{1.75}{0.75}{\omega\hsp}{}}
\put(2,1.5){\Elline{1.75}{0.75}{}{\Hsp\calE-\omega}}
\put(0,4.5){\circle*{0.15}} \put(2,4.5){\circle*{0.15}}
\put(1,0.25){\makebox(0,0){\large\bf{P}}}
\put(1,-0.5){\makebox(0,0){$F(\calE)$}}
\put(5,2.75){\makebox(0,0){\Huge-}}
\end{picture}
\begin{picture}(4,5.5)(-2,0.5)
\put(-0.5,4.5){\dash{12}} \put(-1.3,4.5){\makebox(0,0){$t=t'$}}
\put(-3,2.25){\makebox(0,0){\large$$}}
\put(0,4.5){\Elline{1}{0.5}{\hpsi\dagg}{r}}
\put(2,4.5){\Elline{1}{0.5}{s}{\hpsi\dagg}}
\put(0,3){\Elline{1.5}{0.75}{}{}}
\put(2,3){\Elline{1.5}{0.75}{}{}}
\put(0,1.5){\Elline{1.5}{0.75}{\hpsi}{t}}
\put(2,1.5){\Elline{1.5}{0.75}{u}{\hpsi}} \put(0,3){\photon{}{}{}}
\put(1,1.75){\makebox(0,0){\large\bf{P}}}
\put(3.2,3.25){\makebox(0,0){\Large$\times$}}
\put(1,0.75){\makebox(0,0){$F(\E')$}}
\end{picture}
\begin{picture}(4,5.5)(-2,-1)
\put(0,3){\Elline{1}{0.5}{\hpsi\dagg}{t}}
\put(2,3){\Elline{1}{0.5}{u}{\hpsi\dagg}} \put(-0.5,3){\dash{12}}
\put(3.6,3){\makebox(0,0){$t=0$}}
\put(0,1.5){\Elline{1.5}{0.75}{}{}}
\put(2,1.5){\Elline{1.5}{0.75}{}{}}
\put(0,0){\Elline{1.5}{0.75}{\hpsi}{a}}
\put(2,0){\Elline{1.5}{0.75}{b}{\hpsi}} \put(0,1.5){\photon{}{}{}}
\put(1,0.25){\makebox(0,0){\large\bf{P}}}
\put(1,-0.75){\makebox(0,0){$F(\calE)$}}
\end{picture}
    \renewcommand{\normalsize}{\footnotesize}
    \caption{Graphical representation of the separable
      two-photon-photon ladder diagram (left). This diagram is
      separable, if there exists a time (represented by the
      dotted line) at which there is no uncontracted photon, i.e., a time after the
      first photon has been absorbed and before the second has been created.
      The corresponding counterterm (right) is a product of two operators,
      which evolve independently from
      possibly different states of the model space.}
      \renewcommand{\normalsize}{\standard}
    \label{Fig:SepPh}
  \end{center}
\end{figure}
The counterterm \eq{C} is in the present case given by the product
of two single-photon contributions, as shown in Fig.
\ref{Fig:SepPh} (right)
\begin{equation}
  \label{C1}
  C\tva(t)=-U\ett(t,-\infty)\bsdot PU\ett(0,-\infty)\,P=-U\ett(t)\bsdot
  PU\ett(0)
\end{equation}
using the notation introduced in subsection \ref{sec:MSC}. The two
factors evolve independently from (possibly different) states in
the model space, which is indicated by the "dot". The counterterm
eliminates the singularity, but there may be a \it{finite
remainder}, which we refer to as the model-space contribution
(MSC) \eq{M}. We shall first consider this part.

We assume that we operate to the far right on a function of the
type \eq{FE} of energy $\E$, and that the intermediate model-space
state has the energy $\E'$. Using the first-order result \eq{SPc},
we can express the second-order evolution operator \eq{U2a} as
\begin{equation}
  \label{U2}
  U\tva(t)P=\me^{-\im t(\E-H_0)}\,U\ett(0,\E)\,U\ett(0,\E)P
\end{equation}
and the counterterm---with the first factor evolving from the
intermediate state---as
\begin{equation}
  \label{C2}
  C\tva(t)=-\me^{-\im t(\E'-H_0)}\,U\ett(0,\E')\bsdot PU\ett(0,\E)P
\end{equation}
We note here that the time derivative for $U$ as well as $C$
eliminates the denominator of the leftmost resolvent.

The MSC now becomes
\begin{equation}
  \label{MSC}
  MP=\Big(U\ett(0,\E)-U\ett(0,\E')\Big)\bsdot
  PU\ett(0,\E)P
\end{equation}
Using the result \eq{dots}, the last factor is
  \[PU\ett(0,\E)P=\frac{P}{\E-\E'}\,V(\E)P=-\frac{1}{\deltaE}*PV\ett(\E)P
  =-\frac{1}{\deltaE}*P\Udot\ett(\E)P\]
with  $\deltaE=\E'-\E$, and with
$\;\delta{U\ett}=U\ett(0,\E')-U\ett(0,\E)$ we have
\begin{equation}
  \label{MP1}
  MP=\Big(U\ett(0,\E)-U\ett(0,\E')\Big)\bsdot  PU\ett(0,\E)P
  =\partdelta{U\ett}{\E}*P\Udot\ett P
\end{equation}
(The asterisk is used only for clarity. It notifies the position
of a "fold" in the graphical representation~\cite{LM86}, but has
no other special significance. It will mainly serve as a reminder
of the position of a cancelled singularity, which---as we shall
see---requires certain precautions.) With the definition \eq{Hbar}
the MSC can be expressed
\begin{equation}
  \label{MP}
  MP=\partdelta{U\ett}{\E}*H\eff\ett
\end{equation}
(Note that $\Ubar\ett=U\ett$ and $\Hbar\eff\ett=H\eff\ett$.) The
complete second-order reduced evolution operator \eq{M} then
becomes
\begin{equation}
  \label{MSC1}
  \Util\tva(0)P=\Ubar\tva(0)P+\partdelta{U\ett}{\E}*H\eff\ett
\end{equation}

The result above is exact also for the quasi-degenerate case. The
difference ratio can be expanded as discussed in Appendix
\ref{App:Exp}
\begin{equation}
  \label{MSC2}
  \partdelta{U\ett}{\E}=\partder{U\ett}{\E}+\half\ppartder{U\ett}{\E}\,\deltaE
  +\frac{1}{3!}\pppartder{U\ett}{\E}\,\deltaE^2+\cdots
\end{equation}
which in the limit of complete degeneracy yields
\begin{equation}
  \label{MSC2A}
  \Util\tva(0)P=\Ubar\tva(0)P+\partder{U\ett}{\E}*H\eff\ett
\end{equation}
The second-order contribution to the wave operator \eq{WaveOp}
then becomes
\begin{equation}
  \label{Om2}
  \boxed{\Omega\tva P=Q\Util\tva(0)P
  =\Ombar\tva P+\partdelta{\Omega\ett}{\E}*H\eff\ett
  \Rarr\Ombar\tva P+\partder{\Omega\ett}{\E}*H\eff\ett}
\end{equation}
where $\Ombar$ is the wave operator \eq{Ombar} without
intermediate model space states.

The second-order contribution to the effective interaction is
obtained by means of the relation \eq{EffInt}. Since the
expression \eq{Om2} is valid only for $t=0$, it can not be used to
evaluate the time derivative. Instead, we have to use the original
definition \eq{U+C}, and using the expressions \eq{U2} and
\eq{C2}, we find
\begin{equation}
  \label{Heff22}
  H\eff\tva= PV(\E)\,\G(\E)\,VP-PV(\E')\,\GP(\E)\,V(\E)P
  =\Hbar\tva\eff+\partdelta{H\eff\ett}{\E}*H\eff\ett
\end{equation}
where $\Hbar\tva\eff=PV\GQ VP$ is the $H-bar$ operator \eq{Hbar}
with no intermediate model-space states. The last term is by
definition the model-space contribution, which appears in this
order only for energy-dependent interactions. In the case of
complete degeneracy the difference ratio tends to the derivative,
as before,
\begin{equation}
  \label{Heff2}
  \boxed{H\eff\tva= \Hbar\eff\tva+\partdelta{H\eff\ett}{\E}*H\eff\ett
  \Rarr \Hbar\eff\tva+\partder{H\eff\ett}{\E}*H\eff\ett}
\end{equation}

\subsection{Separable three-photon exchange}
  \label{App:SepThree}
The treatment of the exchange of three separable covariant photons
is quite analogous to the previous case. From the expansion
\eq{OmRec3} we have
\begin{eqnarray}
  \label{U3}
 \Util\tre P&=&\Ubar\tre P
 +\big(\Ubar\tva P\Ubar\ett P-\Ubar\tva P\bsdot\Ubar\ett P\big)
 +\big(\Ubar\ett P\Ubar\tva P-\Ubar\ett P\bsdot\Ubar\tva P\big)\nn
 &+&\big(\Ubar\ett P\Ubar\ett P-\Ubar\ett P\bsdot\Ubar\ett P\big)
  \big(P\Ubar\ett P-\bsdot P\Ubar\ett P\big)
\end{eqnarray}
By generalizing the result of the preceding Appendix we obtain the
relation
\begin{equation}
 \label{GenDiff}
   \boxed{AP\;BP-AP\bsdot B P=\partdelta{A}{\E}*\dot{B}}
\end{equation}
where $A$ is an arbitrary operator and $B$ can be $U$, $\Ubar$ or
$\Util$. Using this relation, the second and third terms above
become
\begin{eqnarray}
 \label{U2PU}
  \Hsp\big(\Ubar\tva P\Ubar\ett P-\Ubar\tva\bsdot P\Ubar\ett
  P\big)=\partdelta{\Ubar\tva}{\E}*P\Ubardot\ett P
  =\partdelta{\Ubar\tva}{\E}*\Hbar\eff\ett\\
  \label{UPU2}
  \Hsp\big(\Ubar\ett P\Ubar\tva P-\Ubar\ett P\bsdot\Ubar\tva P\big)
  =\partdelta{\Ubar\ett}{\E}*P\Ubardot\tva P
  =\partdelta{\Ubar\ett}{\E}*\Hbar\eff\tva
\end{eqnarray}
In the last term in Eq. \eq{U3} we have to apply the rule
\eq{GenDiff} twice, yielding
\begin{eqnarray}
 \label{dUH}
  &&\big(\Ubar\ett P\Ubar\ett P-\Ubar\ett P\bsdot\Ubar\ett P\big)
  \big(P\Ubar\ett P-\bsdot P\Ubar\ett P\big)\nn
   &=&\partdelta{\Ubar\ett}{\E}*P\Ubardot\ett P
  \;\big(P\Ubar\ett P-\bsdot P\Ubar\ett P\big)
  =\partdelta{}{\E}\Big(\partdelta{\Ubar\ett}{\E}*P\Ubardot\ett P\Big)
   *P\Ubardot\ett P\nn
   &=&\partdelta{}{\E}\Big(\partdelta{\Ubar\ett}{\E}*\Hbar\eff\ett\Big)
   *\Hbar\eff\ett
  \end{eqnarray}
From the previous Appendix (Eq. \ref{MSC2}) we have
\begin{equation}
  \label{U2dif}
  \partdelta{\Util\tva}{\E}=\partdelta{\Ubar\tva}{\E}
  +\partdelta{}{\E}\Big(\partdelta{\Ubar\ett}{\E}*\Hbar\ett\eff\Big)
\end{equation}
and the complete result then becomes
\begin{equation}
  \label{U3B}
  \Util\tre P=\Ubar\tre P+\partdelta{\Util\tva}{\E}*H\eff\ett
    +\partdelta{\Ubar\ett}{\E}*\Hbar\eff\tva
\end{equation}
This is an exact expression in this order, also for a
quasi-degenerate model space. In the case of complete degeneracy
this becomes
\begin{equation}
  \label{U3deg}
  \Util\tre P=\Ubar\tre P+\partder{\Util\tva}{\E}*H\eff\ett
    +\partder{\Ubar\ett}{\E}*\Hbar\eff\tva
\end{equation}
In terms of the $\Om$ operators the results above then become
\begin{equation}
  \label{Om3C}
  \boxed{\Omega\tre P=\Ombar\tre P+\partdelta{\Om\tva}{\E}*\Hbar\eff\ett
  +\partdelta{\Omega\ett}{\E}*\Hbar\eff\tva}
\end{equation}

In order to obtain the third-order effective interaction, we
consider the time derivative of the relation \eq{U3} (only the
first factor is time dependent). This yields
\begin{equation}
  H\eff\tre=\Hbar\eff\tre+\partdelta{\Hbar\eff\ett}{\E}*\Hbar\eff\tva
  +\partdelta{\Hbar\eff\tva}{\E}*\H\eff\ett
  +\partdelta{}{\E}\Big(\partdelta{H\ett\eff}{\E}*H\ett\eff\Big)*H\ett\eff
\end{equation}
which using the relation \eq{Heff2} can be expressed
\begin{equation}
  \label{Heff3}
  \boxed{H\eff\tre=\Hbar\eff\tre+\partdelta{H\eff\ett}{\E}*\Hbar\eff\tva
  +\partdelta{H\eff\tva}{\E}*H\eff\ett}
\end{equation}

\subsection{Expansions}
\label{App:Exp} We have seen above that when there are multiple
singularities, it is important to take the difference ratios
\it{before} the singularities are removed. We shall illustrate
this here by a simple mathematical example.

 We consider a function
$f(x)$ of the variable $x$. We define the first-order difference
ratio
\begin{equation}
  \label{Deltaf}
  \partdelta{f}{x}=\partdelta{_{x_0,x}f}{x}=\frac{f(x)-f(x_0)}{x-x_0}
\end{equation}
which can be expanded in a Taylor series
\begin{equation}
  \label{DeltafB}
  \partdelta{f}{x}=\partdelta{_{x_0,x}f}{x}=f'(x_0)+\half f"(x_0)(x-x_0)
  +\frac{1}{3!} f"'(x_0)(x-x_0)^2+\frac{1}{4!} f^{IV}(x_0)(x-x_0)^3+\cdots
\end{equation}
where
\begin{equation}
  \label{f'}
  f'(x_0)=\deriv{f}{x}\Big|_{x=x_0}
\end{equation}
etc.

Similarly, we define the second-order difference ratio
\begin{eqnarray}
  \label{Deltaf2}
  \ppartdelta{f}{x}&=&\partdelta{_{x'x}}{x}\partdelta{_{x_0,x}f}{x}
  =\frac{\partdelta{_{x_0,x'}f}{x}-\partdelta{_{x_0,x}f}{x}}{x'-x}
  =\half f"(x_0)+\frac{1}{3!} f"'(x_0)(x+x'-2x_0)\nn
  &+&\frac{1}{4!} f^{IV}(x_0)\big[(x'-x_0)^2+(x'-x_0)(x-x_0)+(x-x_0)^2\big]+\cdots
\end{eqnarray}
the third-order difference ratio
\begin{equation}
  \label{Deltaf3}
  \pppartdelta{f}{x}=\partdelta{_{x"x'}}{x}\partdelta{_{x'x}}{x}\partdelta{_{x_0,x}f}{x}
  =\frac{1}{3!} f"'(x_0)+\frac{1}{4!} f^{IV}(x_0)(x+x'+x"-3x_0)+\cdots
\end{equation}
the fourth-order difference ratio
\begin{equation}
  \label{Deltaf4}
  \frac{\delta^4f}{\delta^4x}=\frac{1}{4!} f^{IV}(x_0)+\cdots
\end{equation}
and so on.

Generalizing these results, we have in the limit, when the
differences tend to zero
\begin{equation}
  \label{Dif-Der}
  \boxed{\partdelta{^nf}{^nx}\Rarr\frac{1}{n!}\frac{\dif^nf}{\dif^nx}}
\end{equation}
This relation is frequently used in the present paper.

\bibliographystyle{prsty}

\input{Bethe-SalpeterCJP.bbl}
\end{document}

%% file: FigurecommandsV.tex


\newcommand{\LineH}[1]
{\linethickness{0.4mm}
\put(0,0){\line(1,0){#1}}
}
\newcommand{\LineS}[1]
{\linethickness{1mm}
\put(0,0){\line(1,0){#1}}
}

\newcommand{\LineWO}[1]
{\linethickness{0.5mm}
\put(0,0){\line(1,0){#1}}
}

\newcommand{\LineV}[1]
{\linethickness{0.4mm}
\put(0,0){\line(0,1){#1}}
}

\newcommand{\Linev}[1]
{\put(0.,0){\line(0,1){#1}}
}

\newcommand{\LineW}[1]
{\put(0.015,0){\line(0,1){#1}}
\put(0,0){\line(0,1){#1}}
\put(-0.015,0){\line(0,1){#1}}}

\newcommand{\LineHpt}[1]
{\put(0,0.015){\line(1,0){#1}}
\put(0,-0.015){\line(1,0){#1}}
\put(0,0){\circle*{0.15}}
\put(#1,0){\circle*{0.15}}}

\newcommand{\LineDl}[1]
{\put(0.012,-0.012){\line(-1,-1){#1}}\put(-0.012,0.012){\line(-1,-1){#1}}
\put(0.0,0.0){\line(-1,-1){#1}}}

\newcommand{\Linedl}[1]
{\put(0.01,0.01){\line(-1,-2){#1}}
\put(-0.01,-0.01){\line(-1,-2){#1}}}

\newcommand{\LineUr}[1]
{\put(0,0.015){\line(1,1){#1}}
\put(0,0){\line(1,1){#1}}
\put(0,-0.015){\line(1,1){#1}}}

\newcommand{\LineDr}[1]
{\put(0,0.01){\line(1,-1){#1}} \put(0,0){\line(1,-1){#1}}
\put(0,-0.01){\line(1,-1){#1}}}

\newcommand{\Linedr}[1]
{\put(0,0.01){\line(1,-3){#1}}
\put(0,0){\line(1,-3){#1}}
\put(0,-0.01){\line(1,-3){#1}}}

\newcommand{\Lineur}[1]
{\put(0,0.01){\line(1,3){#1}}
\put(0,0){\line(1,3){#1}}
\put(0,-0.01){\line(1,3){#1}}}

\newcommand{\DLine}[1]
{\put(0.,-0.05){\line(1,0){#1}}
\put(0.,0.05){\line(1,0){#1}}
\put(0,0){\circle*{0.1}}}

\newcommand{\Vector}[0]
{\thicklines\setlength{\unitlength}{1cm}\put(-0.13,0){\vector(-1,0){0}}}

\newcommand{\VectorR}[0]
{\thicklines\put(0.13,0.0){\vector(1,0){0}}}

\newcommand{\VectorUp}[0]
{\thicklines\setlength{\unitlength}{1cm}
\put(-0,0.12){\vector(0,0){0}}}

\newcommand{\VectorDn}[0]
{\thicklines\setlength{\unitlength}{1cm}
\put(0.012,-0.12){\vector(0,-1){0}}
\put(-0.012,-0.12){\vector(0,-1){0}}}

\newcommand{\VectorDl}[0]
{\thicklines
\setlength{\unitlength}{1cm}
\put(-0.092,-0.076){\vector(-1,-1){0}}
\put(-0.076,-0.092){\vector(-1,-1){0}}
}

\newcommand{\VectorDr}[0]
{\thicklines
\setlength{\unitlength}{1cm}
\put(0.076,-0.092){\vector(1,-1){0}}
\put(0.092,-0.076){\vector(1,-1){0}}
}

\newcommand{\Vectordr}[0]
{\setlength{\unitlength}{1cm}
\put(0.022,0.112){\vector(1,-3){0}}
\put(0.052,0.118){\vector(1,-3){0}}
}

\newcommand{\Vectorur}[0]
{\setlength{\unitlength}{1cm}
\put(0.04,-0.062){\vector(1,3){0}}
\put(0.06,-0.068){\vector(1,3){0}}
}

\newcommand{\VectorUr}[0]
{\put(0.195,0.705){\vector(1,1){0}}
 \put(0.22,0.76){\vector(1,1){0}}
\put(0.155,0.735){\vector(1,1){0}} }

\newcommand{\VectorUl}[0]
{\put(-0.23,-0.02){\vector(-1,1){0}}
\put(-0.19,-0.03){\vector(-1,1){0}}
\put(-0.22,-0.06){\vector(-1,1){0}}
}

\newcommand{\Wector}[0]
{\put(-0.15,0)\Vector\put(0.15,0)\Vector}

\newcommand{\WectorUp}[0]
{\put(0,0.125)\VectorUp\put(0,-0.125)\VectorUp}

\newcommand{\WectorDn}[0]
{\put(0,0.125)\VectorDn\put(0,-0.125)\VectorDn}

\newcommand{\WectorDl}[0]
{\put(0.1,0.1)\VectorDl\put(-0.1,-0.1)\VectorDl}

\newcommand{\EllineH}[4]
{\put(0,0){\LineH{#1}}
\put(#2,0){\Vector}
\put(#2,0.45){\makebox(0,0){$#3$}}
\put(#2,-0.35){\makebox(0,0){$#4$}}}

\newcommand{\lline}[4]
{\put(0,0){\LineV{#1}} \put(-0.3,#2){\makebox(0,0){$#3$}}
\put(0.4,#2){\makebox(0,0){$#4$}}}

\newcommand{\Elline}[4]
{\put(0,0){\LineV{#1}} \put(0,#2){\VectorUp}
\put(-0.3,#2){\makebox(0,0){$#3$}}
\put(0.4,#2){\makebox(0,0){$#4$}}}

\newcommand{\DElline}[4]
{\put(0,0){\LineV{#1}}
\put(0,#2){\WectorUp}
\put(-0.3,#2){\makebox(0,0){$#3$}}
\put(0.3,#2){\makebox(0,0){$#4$}}}

\newcommand{\DEllineDn}[4]
{\put(0,0){\LineV{#1}}
\put(0,#2){\WectorDn}
\put(-0.25,#2){\makebox(0,0){$#3$}}
\put(0.25,#2){\makebox(0,0){$#4$}}}

\newcommand{\Ellinet}[4]
{\put(0,0){\Linev{#1}} \put(0,#2){\vector(0,1){0}}
\put(-0.35,#2){\makebox(0,0){$#3$}}
\put(0.35,#2){\makebox(0,0){$#4$}}}
\newcommand{\EllineT}[4]
{\put(0,0){\LineW{#1}}
\put(0,#2){\VectorUp}
\put(-0.35,#2){\makebox(0,0){$#3$}}
\put(0.35,#2){\makebox(0,0){$#4$}}}

\newcommand{\EllineDnt}[4]
{\put(0,0){\Linev{#1}}
\put(0,#2){\VectorDn}
\put(-0.35,#2){\makebox(0,0){$#3$}}
\put(0.35,#2){\makebox(0,0){$#4$}}}

\newcommand{\EllineDn}[4]
{\put(0,0){\LineV{#1}} \put(0,#2){\VectorDn}
\put(-0.35,#2){\makebox(0,0){$#3$}}
\put(0.35,#2){\makebox(0,0){$#4$}}}

\newcommand{\EllineDl}[4]
{\put(0,0){\LineDl{#1}} \put(-#2,-#2){\VectorDl}
\put(-0.2,0.4){\makebox(-#1,-#1){$#3$}}
\put(0.2,-0.2){\makebox(-#1,-#1){$#4$}}}

\newcommand{\Ellinedl}[4]
{\put(0.01,0.01){\line(-1,-2){#1}}
\put(-0.01,-0.01){\line(-1,-2){#1}}
\thicklines\put(-0.05,-0.1){\vector(-1,-2){#2}}
\put(-0.2,0.2){\makebox(-#1,-#1){$#3$}}
\put(0.2,-0.2){\makebox(-#1,-#1){$#4$}}}

\newcommand{\Ellinedr}[4]
{\put(0.01,0.01){\line(1,-2){#1}}
\put(-0.01,-0.01){\line(1,-2){#1}}
\thicklines\put(0,0){\vector(1,-2){#2}}
\put(0.2,-0.3){\makebox(#1,-#1){$#3$}}
\put(0,-1){\makebox(#1,-#1){$#4$}}}

\newcommand{\EllineA}[7]
{\put(0.0,0.0){\line(#1,#2){#3}}
\put(0.005,0.0){\line(#1,#2){#3}}
\put(-0.005,0.0){\line(#1,#2){#3}}
\put(0,0){\vector(#1,#2){#4}}
\put(0.010,0){\vector(#1,#2){#4}}
\put(-0.010,0){\vector(#1,#2){#4}}
\put(#6,#7){\makebox(0,0){$#5$}}}

\newcommand{\EllineDr}[4]
{\put(0,0){\LineDr{#1}} \put(#2,-#2){\VectorDr}
\put(#2,-#2){\makebox(-0.5,-0.5){$#3$}}
\put(#2,-#2){\makebox(0.5,0.5){$#4$}}}

\newcommand{\EllinedR}[5]
{\put(0,0){\line(1,-3){#1}}
\put(0.014,0){\line(1,-3){#1}}
\put(-0.014,0){\line(1,-3){#1}}
\put(#2,-#3){\makebox(0,0){{\Vectordr}}}
\put(#2,-#3){\makebox(-0.5,-0.5){$#4$}}
\put(#2,-#3){\makebox(0.5,0.5){$#5$}}}

\newcommand{\EllineuR}[5]
{\put(0,0){\line(1,3){#1}}
\put(0.014,0){\line(1,3){#1}}
\put(-0.014,0){\line(1,3){#1}}
\put(#2,#3){\makebox(0,0){\Vectorur}}
\put(#2,#3){\makebox(-0.5,-0.5){$#4$}}
\put(#2,#3){\makebox(0.5,0.5){$#5$}}}


\newcommand{\Ellineur}[5]
{\put(0,0){\Lineur{#1}}
\put(#2,#3){\makebox(0,0){\Vectorur}}
\put(#2,#3){\makebox(-0.5,0){$#4$}}
\put(#2,#3){\makebox(0.5,0){$#5$}}}

\newcommand{\EllineUr}[4]
{\put(0,0){\LineUr{#1}}
\put(#2,#2){\makebox(-0.35,-0.35){\VectorUr}}
\put(-0.2,0.4){\makebox(#1,#1){$#3$}}
\put(0.2,-0.3){\makebox(#1,#1){$#4$}}}

\newcommand{\DEllineDl}[4]
{\put(0,0){\LineDl{#1}}
\put(-#2,-#2){\WectorDl}
\put(-0.25,0.25){\makebox(-#1,-#1){$#3$}}
\put(0.25,-0.25){\makebox(-#1,-#1){$#4$}}}

\newcommand{\Ebox}[2]
{\put(0,0){\LineH{#1}}
\put(0,#2){\LineH{#1}}
\put(0,0){\LineV{#2}}
\put(#1,0){\LineV{#2}}}

\newcommand{\dashH}
{\multiput(0.05,0)(0.25,0){5}{\line(1,0){0.15}}}

\newcommand{\dash}[1]
{\multiput(0.05,0)(0.25,0){#1}{\line(1,0){0.15}}}

\newcommand{\dashV}[1]
{\multiput(0.05,0)(0,0.25){#1}{\line(0,1){0.15}}}

\newcommand{\dashHp}
{\multiput(0.05,0)(0.25,0){6}{\line(1,0){0.15}}}

\newcommand{\DashH}
{\multiput(0.05,0)(0.25,0){10}{\line(1,0){0.15}}}

\newcommand{\dashHnum}[2]
{\multiput(0.05,0)(0.25,0){5}{\line(1,0){0.15}}
\put(-0.25,0){\makebox(0,0){$#1$}}
\put(1.5,0){\makebox(0,0){$#2$}}}

\newcommand{\dashHnuma}[2]
{\multiput(0.05,0)(0.25,0){5}{\line(1,0){0.15}}
\put(0.25,0.25){\makebox(0,0){$#1$}}
\put(1,0.25){\makebox(0,0){$#2$}}}

\newcommand{\dashHnumu}[2]
{\multiput(0.05,0)(0.25,0){5}{\line(1,0){0.15}}
\put(0.25,-0.25){\makebox(0,0){$#1$}}
\put(1,-0.25){\makebox(0,0){$#2$}}}

\newcommand{\Potint}
{\put(0,0)\dashH \put(1.35,0){\makebox(0,0){x}}
\put(0,0){\circle*{0.15}}}

\newcommand{\potint}
{\multiput(0.05,0)(0.25,0){3}{\line(1,0){0.15}}
\put(0.85,0){\makebox(0,0){x}}
\put(0,0){\circle*{0.15}}}

\newcommand{\PotintS}
{\put(0,0){\dash{4}} \put(1,0){\makebox(0,0){$\times$}}
\put(0,0){\circle*{0.1}}}

\newcommand{\PotintL}
{\put(-1.25,0)\dashH
\put(-1.35,0){\makebox(0,0){x}}
\put(0,0){\circle*{0.15}}}

\newcommand{\Effpot}
{\put(0,0)\dashH
\put(1.35,0){\makebox(0,0){x}}
\put(1.35,0){\circle{0.3}}
\put(0,0){\circle*{0.15}}}

\newcommand{\effpot}
{\multiput(0.05,0)(0.25,0){3}{\line(1,0){0.15}}
\put(0.85,0){\makebox(0,0){x}}
\put(0.85,0){\circle{0.3}}
\put(0,0){\circle*{0.1}}}

\newcommand{\EffpotL}
{\put(-1.25,0)\dashH
\put(0,0){\makebox(0,0){x}}
\put(-1.25,0){\circle{0.3}}
\put(-1.25,0){\circle*{0.15}}}

\newcommand{\Triang}
{\put(0,0){\line(2,1){0.5}}
\put(0,0){\line(2,-1){0.5}}
\put(0.5,-0.25){\line(0,1){0.5}}}

\newcommand{\TriangL}
{\put(0,0){\line(-2,1){0.5}}
\put(0,0){\line(-2,-1){0.5}}
\put(-0.5,-0.25){\line(0,1){0.5}}}

\newcommand{\hfint}
{\put(0,0)\dashH
\put(1.25,0){\makebox(0,0){\Triang}}
\put(0,0){\circle*{0.15}}}

\newcommand{\hfintL}
{\put(-1.25,0)\dashH
\put(-1.25,0){\makebox(0,0){\TriangL}}
\put(0,0){\circle*{0.15}}}

\newcommand{\VPloop}[1]
{\put(0,0){\circle{1}}
\put(0,0.0){\circle{1.}}
\put(0.02,0){\circle{1.}}
\put(0,0.02){\circle{1.}}
\put(0,-0.02){\circle{1}}
\put(-0.02,0){\circle{1}}
\put(0.52,0.05){\VectorDn}
\put(0.85,0){\makebox(0,0){$#1$}}}

\newcommand{\VPloopt}[1]
{\put(0,0){\circle{1}}
\put(0.52,0.05){\VectorDn}
\put(0.75,0){\makebox(0,0){$#1$}}}

\newcommand{\VPloopL}[1]
{\put(0,0){\circle{1}}
\put(0.01,0.){\circle{1}}\put(-0.01,0){\circle{1}}
\put(0,0.01){\circle{1}}\put(0,-0.01){\circle{1}}
\put(-0.5,0.05){\VectorDn}
\put(-0.75,0){\makebox(0,0){$#1$}}}

\newcommand{\VPloopLt}[1]
{\put(0,0){\circle{1}}
\put(-0.5,0){\VectorDn}
\put(-0.75,0){\makebox(0,0){$#1$}}}

\newcommand{\VPloopLR}[2]
{\put(0,0){\circle{1}}\put(-0.01,0){\circle{1}}
\put(0,0.01){\circle{1}}\put(0,-0.01){\circle{1}}
\put(-0.5,0){\VectorDn}
\put(0.5,0){\VectorUp}
\put(-0.75,0){\makebox(0,0){$#1$}}
\put(0.75,0){\makebox(0,0){$#2$}}}

\newcommand{\VPloopLRt}[2]
{\put(0,0){\circle{1}}
\put(-0.5,0){\VectorDn}
\put(0.5,0){\VectorUp}
\put(-0.75,0){\makebox(0,0){$#1$}}
\put(0.75,0){\makebox(0,0){$#2$}}}

\newcommand{\VPloopD}[2]
{\put(0,0){\circle{1}}
\put(0.01,0.){\circle{1}}\put(-0.01,0){\circle{1}}
\put(0.,0.){\circle{1}}\put(-0.01,0){\circle{1}}
\put(0,0.01){\circle{1}}\put(0,-0.01){\circle{1}}
\put(0,0.52){\VectorR}
\put(0,-0.52){\Vector}
\put(0,0.8){\makebox(0,0){$#1$}}
\put(0,-0.8){\makebox(0,0){$#2$}}}

\newcommand{\VPloopDt}[2]
{\put(0,0){\circle{1}}
\put(0,0.5){\VectorR}
\put(0.05,-0.47){\Vector}
\put(0,0.8){\makebox(0,0){$#1$}}
\put(0,-0.8){\makebox(0,0){$#2$}}}

\newcommand{\Loop}[2]
{\put(0,0){\oval(0.6,1.25)}\put(0.01,0.01){\oval(0.6,1.25)}\put(-0.01,-0.01){\oval(0.6,1.25)}
\put(0.3,0){\VectorUp}
\put(-0.3,0){\VectorDn}
\put(-0.65,0){\makebox(0,0){$#1$}}
\put(0.65,0){\makebox(0,0){$#2$}}}

\newcommand{\HFexch}[1]
{\put(0,0)\dashH
\qbezier(0,0.01)(0.625,0.515)(1.25,0.015)
\qbezier(0,-0.01)(0.625,0.485)(1.25,-0.015)
\put(0.625,0.26){\Vector}
\put(0.625,0.5){\makebox(0,0){$#1$}}
\put(0,0){\circle*{0.15}}
\put(1.25,0){\circle*{0.15}}}

\newcommand{\HFexcht}[1]
{\put(0,0)\dashH
\qbezier(0,0.01)(0.625,0.515)(1.25,0.015)
\put(0.625,0.26){\Vector}
\put(0.625,0.5){\makebox(0,0){$#1$}}
\put(0,0){\circle*{0.15}}
\put(1.25,0){\circle*{0.15}}}

\newcommand{\Dashpt}[1]
{\multiput(0,-0.6)(0,0.25){13}{\line(0,1){0.15}}
\put(0,0){\circle*{0.15}}
\put(0,#1){\circle*{0.15}}}

\newcommand{\photonH}[3]
{\qbezier(0,0)(0.08333,0.125)(0.1666667,0)
\qbezier(0.1666667,0)(0.25,-0.125)(0.3333333,0)
\qbezier(0.3333333,0)(0.416667,0.125)(0.5,0)
\qbezier(0.5,0)(0.583333,-0.125)(0.666667,0)
\qbezier(0.666667,0)(0.75,0.125)(0.833333,0)
\qbezier(0.833333,0)(0.916667,-0.125)(1,0)
\qbezier(1,0)(1.083333,0.125)(1.166667,0)
\qbezier(1.166667,0)(1.25,-0.125)(1.333333,0)
\qbezier(1.333333,0)(1.416667,0.125)(1.5,0)
\put(0.75,0.){\VectorR} \put(0.75,0.35){\makebox(0,0){$#1$}}
\put(0,0){\circle*{0.1}} \put(1.5,0){\circle*{0.1}}
\put(-0.5,0){\makebox(0,0){#2}} \put(2,0){\makebox(0,0){#3}}}

\newcommand{\photon}[3]
{\qbezier(0,0)(0.08333,0.125)(0.1666667,0)
\qbezier(0.1666667,0)(0.25,-0.125)(0.3333333,0)
\qbezier(0.3333333,0)(0.416667,0.125)(0.5,0)

\qbezier(0.5,0)(0.583333,-0.125)(0.666667,0)
\qbezier(0.666667,0)(0.75,0.125)(0.833333,0)
\qbezier(0.833333,0)(0.916667,-0.125)(1,0)

\qbezier(1,0)(1.083333,0.125)(1.166667,0)
\qbezier(1.166667,0)(1.25,-0.125)(1.333333,0)
\qbezier(1.333333,0)(1.416667,0.125)(1.5,0)

\qbezier(1.5,0)(1.583333,-0.125)(1.666667,0)
\qbezier(1.666667,0)(1.75,0.125)(1.833333,0)
\qbezier(1.833333,0)(1.916667,-0.125)(2,0)
\put(1,0.0){\VectorR}
\put(1,0.35){\makebox(0,0){$#1$}}
\put(0,0){\circle*{0.15}}
\put(2,0){\circle*{0.15}}
\put(-0.35,0){\makebox(0,0){#2}}
\put(2.35,0){\makebox(0,0){#3}}}

\newcommand{\photonHS}[4]
{\qbezier(0,0)(0.08333,0.125)(0.1666667,0)
\qbezier(0.1666667,0)(0.25,-0.125)(0.3333333,0)
\qbezier(0.3333333,0)(0.416667,0.125)(0.5,0)
\qbezier(0.5,0)(0.583333,-0.125)(0.666667,0)
\qbezier(0.666667,0)(0.75,0.125)(0.833333,0)
\qbezier(0.833333,0)(0.916667,-0.125)(1,0)
\put(0.5,0.025){\VectorR}
\put(0.5,0.35){\makebox(0,0){$#1$}}
\put(0,0){\circle*{0.15}}
\put(1,0){\circle*{0.15}}
\put(0,-0.5){\makebox(0,0){#2}}
\put(1,-0.5){\makebox(0,0){#3}}}

\newcommand{\photonNE}[3]
{\qbezier(0,0)(0.22,-0.02)(0.2,0.2)
\qbezier(0.2,0.2)(0.18,0.42)(0.4,0.4)
\qbezier(0.4,0.4)(0.62,0.38)(0.6,0.6)
\qbezier(0.6,0.6)(0.58,0.82)(0.8,0.8)
\qbezier(0.8,0.8)(1.02,0.78)(1,1)
\qbezier(1,1)(0.98,1.22)(1.2,1.2)
\qbezier(1.2,1.2)(1.42,1.18)(1.4,1.4)
\qbezier(1.4,1.4)(1.38,1.62)(1.6,1.6)
\qbezier(1.6,1.6)(1.82,1.58)(1.8,1.8)
\qbezier(1.8,1.8)(1.78,2.02)(2,2)
\put(1,1){\makebox(0.05,-0.2){\VectorUp}} \put(0,0){\circle*{0.1}}
\put(2,2){\circle*{0.1}} \put(1,1){\makebox(-0.6,0.4){$#1$}}
\put(-0.35,-1){\makebox(0,2){$#2$}}
\put(2.35,1){\makebox(0,2){$#3$}}}

\newcommand{\photonNNE}[3]
{\qbezier(0,0)   (0.28,-0.02)(0.2,0.3)
\qbezier(0.2,0.3)(0.12,0.52)(0.4,0.6)
\qbezier(0.4,0.6)(0.68,0.58)(0.6,0.9)
\qbezier(0.6,0.9)(0.52,1.12)(0.8,1.2)
\qbezier(0.8,1.2)(1.08,1.18)(1,1.5) \qbezier(1,1.5)
(0.92,1.72)(1.2,1.8) \qbezier(1.2,1.8)(1.48,1.86)(1.4,2.1)
\qbezier(1.4,2.1)(1.365,2.24)(1.6,2.4)
\qbezier(1.6,2.4)(1.835,2.46)(1.8,2.7)
\qbezier(1.8,2.7)(1.765,2.84)(2,3)
\put(0.6,0.8){\makebox(0,0){\VectorUp}} \put(0,0){\circle*{0.1}}
\put(2,3){\circle*{0.1}} \put(1,0.8){\makebox(0,0){$#1$}}
\put(-0.35,0){\makebox(0,0){$#2$}}
\put(2.35,3){\makebox(0,0){$#3$}}}

\newcommand{\photonENE}[3]
{\qbezier(0,0)(0.17,-0.04)(0.2,0.1)
\qbezier(0.2,0.1)(0.23,0.32)(0.4,0.2)
\qbezier(0.4,0.2)(0.57,0.16)(0.6,0.3)
\qbezier(0.6,0.3)(0.63,0.52)(0.8,0.4)
\qbezier(0.8,0.4)(0.97,0.36)(1,0.5)
\qbezier(1,0.5)(1.03,0.72)(1.2,0.6)
\qbezier(1.2,0.6)(1.37,0.56)(1.4,0.7)
\qbezier(1.4,0.7)(1.43,0.92)(1.6,0.8)
\qbezier(1.6,0.8)(1.77,0.76)(1.8,0.9)
\qbezier(1.8,0.9)(1.83,1.12)(2,1)
\put(1.2,0.75){\makebox(-0.1,0.06){\VectorR}}
\put(1.2,0.85){\makebox(0,-0){$#1$}}
\put(-0.35,-1){\makebox(0,2){$#2$}}
\put(2.35,0){\makebox(0,2){$#3$}}}

\newcommand{\photonNW}[3]
{\qbezier(0,0)(-0.22,-0.02)(-0.2,0.2)
\qbezier(-0.2,0.2)(-0.18,0.42)(-0.4,0.4)
\qbezier(-0.4,0.4)(-0.62,0.38)(-0.6,0.6)
\qbezier(-0.6,0.6)(-0.58,0.82)(-0.8,0.8)
\qbezier(-0.8,0.8)(-1.02,0.78)(-1,1) \qbezier(-1,1)
(-0.98,1.22)(-1.2,1.2) \qbezier(-1.2,1.2)(-1.42,1.18)(-1.4,1.4)
\qbezier(-1.4,1.4)(-1.38,1.62)(-1.6,1.6)
\qbezier(-1.6,1.6)(-1.82,1.58)(-1.8,1.8)
\qbezier(-1.8,1.8)(-1.78,2.02)(-2,2)
\put(-1,1){\makebox(0,-0.2){\VectorUp}} \put(0,0){\circle*{0.1}}
\put(-2,2){\circle*{0.1}} \put(-1,1){\makebox(0.4,0.7){$#1$}}
\put(-2.35,2){\makebox(0,0){$#2$}}
\put(0.35,0){\makebox(0,0){$#3$}}}

\newcommand{\photonSEst}[3]
{\qbezier(0,0)(-0.22,-0.02)(-0.2,0.2)
\qbezier(-0.2,0.2)(-0.18,0.42)(-0.4,0.4)
\qbezier(-0.4,0.4)(-0.62,0.38)(-0.6,0.6)
\qbezier(-0.6,0.6)(-0.58,0.82)(-0.8,0.8)
\qbezier(-0.8,0.8)(-1.02,0.78)(-1,1) \qbezier(-1,1)
(-0.98,1.22)(-1.2,1.2) \qbezier(-1.2,1.2)(-1.42,1.18)(-1.4,1.4)
\qbezier(-1.4,1.4)(-1.38,1.62)(-1.6,1.6)
\qbezier(-1.6,1.6)(-1.82,1.58)(-1.8,1.8)
\qbezier(-1.8,1.8)(-1.78,2.02)(-2,2)
\put(-1,1){\makebox(0,0){\VectorDn}} \put(0,0){\circle*{0.1}}
\put(-2,2){\circle*{0.1}} \put(-1,1){\makebox(0.4,0.7){$#1$}}
\put(-2.35,2){\makebox(0,0){$#2$}}
\put(0.35,0){\makebox(0,0){$#3$}}}

\newcommand{\photonWNW}[3]
{\qbezier(0,0)(-0.17,-0.04)(-0.2,0.1)
\qbezier(-0.2,0.1)(-0.23,0.32)(-0.4,0.2)
\qbezier(-0.4,0.2)(-0.57,0.16)(-0.6,0.3)
\qbezier(-0.6,0.3)(-0.63,0.52)(-0.8,0.4)
\qbezier(-0.8,0.4)(-0.97,0.36)(-1,0.5)
\qbezier(-1,0.5)(-1.03,0.72)(-1.2,0.6)
\qbezier(-1.2,0.6)(-1.37,0.56)(-1.4,0.7)
\qbezier(-1.4,0.7)(-1.43,0.92)(-1.6,0.8)
\qbezier(-1.6,0.8)(-1.77,0.76)(-1.8,0.9)
\qbezier(-1.8,0.9)(-1.83,1.12)(-2,1)
\put(-1,1){\makebox(-0.2,-0.4){$\;$\VectorR}}
\put(0,0){\circle*{0.1}} \put(-2,1){\circle*{0.1}}
\put(-1,0.5){\makebox(0.6,0.4){$#1$}}
\put(0.35,-1){\makebox(0,2){$#3$}}
\put(-2.35,0){\makebox(0,2){$#2$}}}

\newcommand{\Crossphotons}[6]
{\put(0,0){\photonNE{#5}{}{}}
\put(2,0){\photonNW{#6}{}{}}
\put(-0.35,0){\makebox(0,0){$#1$}}
\put(2.35,2){\makebox(0,0){$#2$}}
\put(2.35,0){\makebox(0,0){$#3$}}
\put(-0.35,2){\makebox(0,0){$#4$}}
}

\newcommand{\photonNe}[3]
{\qbezier(0,0)(0.22,-0.02)(0.2,0.2)
\qbezier(0.2,0.2)(0.18,0.42)(0.4,0.4)
\qbezier(0.4,0.4)(0.62,0.38)(0.6,0.6)
\qbezier(0.6,0.6)(0.58,0.82)(0.8,0.8)
\qbezier(0.8,0.8)(1.02,0.78)(1,1)
\qbezier(1,1)(0.98,1.22)(1.2,1.2)
\qbezier(1.2,1.2)(1.42,1.18)(1.4,1.4)
\qbezier(1.4,1.4)(1.38,1.5)(1.5,1.5)
\put(0.9,1.02){\makebox(0,0.01){\VectorR}}
\put(0.75,0.75){\makebox(-0.6,0.4){$#1$}}
\put(-0.35,0){\makebox(0,0){$#2$}}
\put(1.85,1.5){\makebox(0,0){$#3$}}}

\newcommand{\photonNw}[3]
{\qbezier(0,0)(0.02,0.22)(-0.2,0.2)
\qbezier(-0.2,0.2)(-0.42,0.18)(-0.4,0.4)
\qbezier(-0.4,0.4)(-0.38,0.62)(-0.6,0.6)
\qbezier(-0.6,0.6)(-0.82,0.58)(-0.8,0.8)
\qbezier(-0.8,0.8)(-0.78,1.02)(-1,1) \qbezier(-1,1)
(-1.22,0.98)(-1.2,1.2) \qbezier(-1.2,1.2)(-1.18,1.42)(-1.4,1.4)
\qbezier(-1.4,1.4)(-1.5,1.38)(-1.5,1.5)
\put(-0.5,0.8){\makebox(0,-0){\VectorR}}
\put(-0.7,0.7){\makebox(1,0.3){$#1$}}
\put(0.35,0){\makebox(0,0){$#3$}}
\put(-1.85,1.5){\makebox(0,0){$#2$}} }

\newcommand{\elstat}[3]
{\multiput(0.06,0)(0.25,0){8}{\line(1,0){0.15}}
\put(1,0.35){\makebox(0,0){$#1$}} 
\put(-0.35,0){\makebox(0,0){#2}}
\put(2.35,0){\makebox(0,0){#3}}}

\newcommand{\Multiline}[3]
{\linethickness{0.2mm} \put(0,-0.1){\line(1,0){#1}}
 \put(0,0){\line(1,0){#1}}\put(0,0.1){\line(1,0){#1}}
 \put(-0.35,0){\makebox(0,0){#2}}
\put(2.35,0){\makebox(0,0){#3}}}

\newcommand{\elstatH}[3]
{\multiput(0.06,0)(0.25,0){6}{\line(1,0){0.15}}
\put(0.75,0.25){\makebox(0,0){$#1$}} 
\put(-0.35,0){\makebox(0,0){#2}}
\put(2.35,0){\makebox(0,0){#3}}}

\newcommand{\BreitH}[3]
{\multiput(0.15,0)(0.3,0){5}{\circle*{0.1}}
\put(0.75,0.25){\makebox(0,0){$#1$}} 
\put(-0.35,0){\makebox(0,0){#2}}
\put(2.35,0){\makebox(0,0){#3}}}

\newcommand{\RetBreitDH}[3]
{\small\multiput(0.15,0.3)(0.3,-0.15){5}{\circle*{0.1}}
\put(0.75,0.25){\makebox(0,0){$#1$}}
\put(-0.35,0){\makebox(0,0){#2}} \put(2.35,0){\makebox(0,0){#3}}}

\newcommand{\RetBreitH}[3]
{\small\multiput(0.15,-0.3)(0.3,0.15){5}{\circle*{0.1}}
\put(0.75,0.25){\makebox(0,0){$#1$}}
\put(-0.35,0){\makebox(0,0){#2}} \put(2.35,0){\makebox(0,0){#3}}}

\newcommand{\elsta}[3]
{\multiput(0.06,0)(0.25,0){4}{\line(1,0){0.15}}
\put(1,0.35){\makebox(0,0){$#1$}}
\put(0,0){\circle*{0.1}}
\put(1,0){\circle*{0.1}}
\put(-0.35,0){\makebox(0,0){#2}}
\put(1.35,0){\makebox(0,0){#3}}}

\newcommand{\elstatNO}[3]
{\multiput(0.06,0.08)(0.01,0.01){14}{\tiny.}
\multiput(0.30,0.32)(0.01,0.01){14}{\tiny.}
\multiput(0.55,0.57)(0.01,0.01){14}{\tiny.}
\multiput(0.79,0.81)(0.01,0.01){14}{\tiny.}
\multiput(1.03,1.05)(0.01,0.01){14}{\tiny.}
\multiput(1.27,1.29)(0.01,0.01){14}{\tiny.}
\multiput(1.51,1.53)(0.01,0.01){14}{\tiny.}
\multiput(1.75,1.77)(0.01,0.01){14}{\tiny.}
\put(0.95,0.75){\makebox(0,0){\VectorUr}}
\put(0,0){\circle*{0.15}}
\put(2,2){\circle*{0.15}}
\put(0.75,1.25){\makebox(0,0){$#1$}}
\put(-0.5,0){\makebox(0,0){$#2$}}
\put(2.5,2){\makebox(0,0){$#3$}}
}

\newcommand{\elstatNW}[3]
{\multiput(-0.05,0.05)(-0.015,0.015){10}{\circle*{0.02}}
\multiput(-0.3,0.3)(-0.015,0.015){10}{\circle*{0.02}}
\multiput(-0.55,0.55)(-0.015,0.015){10}{\circle*{0.02}}
\multiput(-0.8,0.8)(-0.015,0.015){10}{\circle*{0.03}}
\multiput(-1.05,1.05)(-0.015,0.015){10}{\circle*{0.03}}
\multiput(-1.3,1.3)(-0.015,0.015){10}{\circle*{0.03}}
\multiput(-1.55,1.55)(-0.015,0.015){10}{\circle*{0.03}}
\multiput(-1.8,1.8)(-0.015,0.015){10}{\circle*{0.03}}
\put(-0.9,0.9){\makebox(0,0){\VectorUl}}
\put(0,0){\circle*{0.215}}
\put(-2,2){\circle*{0.2153}}
\put(-0.9,0.9){\makebox(0,0){\VectorUl}}
\put(0,0){\circle*{0.215}}
\put(-2,2){\circle*{0.215}}
\put(-0.5,1){\makebox(0,0){$#1$}}
\put(0,-0.5){\makebox(0,0){$#2$}}
\put(-2,2.5){\makebox(0,0){$#3$}}
}

\newcommand{\photonSE}[5]
{\put(0,0){\photonHS{#3}{#4}{}{}}
\put(1.5,0){\VPloopD{#1}{#2}}
\put(2,0){\photonHS{#5}{}{}{}}}

\newcommand{\photonSEt}[5]
{\put(0,0){\photonHS{#3}{#4}{}{}}
\put(1.5,0){\VPloopDt{#1}{#2}}
\put(2,0){\photonHS{#5}{}{}{}}}

\newcommand{\ElSE}[3]
{\qbezier(0,-1)(.2025,-1.1489)(0.3420,-0.9397)
\qbezier(0.3420,-0.9397)(0.4167,-0.7217)(0.6428,-0.766)
\qbezier(0.6428,-0.766)(0.8937,-0.7499)(0.866,-0.5)
\qbezier(0.866,-0.5)(0.7831,-0.2850)(0.9848,-0.1736)
\qbezier(0.9848,-0.1736)(1.1667,0)(0.9848,0.1736)
\qbezier(0.9848,0.1736)(0.7831,0.2850)(0.866,0.5)
\qbezier(0.866,0.5)(0.8937,0.7499)(0.6428,0.766)
\qbezier(0.6428,0.766)(0.4167,0.7217)(0.3420,0.9397)
\qbezier(0.3420,0.9397)(.2025,1.1489)(0,1)
\put(1.05,0.02){\VectorUp}
\put(0,1){\circle*{0.15}}
\put(0,-1){\circle*{0.15}}
\put(1.45,0){\makebox(0,0){$#1$}}
\put(-0.35,-1){\makebox(0,0){#2}}
\put(-0.35,1){\makebox(0,0){#3}}}

\newcommand{\ElSEL}[3]
{\qbezier(0,-1)(-.2025,-1.1489)(-0.3420,-0.9397)
\qbezier(-0.3420,-0.9397)(-0.4167,-0.7217)(-0.6428,-0.766)
\qbezier(-0.6428,-0.766)(-0.8937,-0.7499)(-0.866,-0.5)
\qbezier(-0.866,-0.5)(-0.7831,-0.2850)(-0.9848,-0.1736)
\qbezier(-0.9848,-0.1736)(-1.1667,0)(-0.9848,0.1736)
\qbezier(-0.9848,0.1736)(-0.7831,0.2850)(-0.866,0.5)
\qbezier(-0.866,0.5)(-0.8937,0.7499)(-0.6428,0.766)
\qbezier(-0.6428,0.766)(-0.4167,0.7217)(-0.3420,0.9397)
\qbezier(-0.3420,0.9397)(-.2025,1.1489)(0,1)
\put(-1,0.02){\VectorUp} 
\put(-1.45,0){\makebox(0,0){$#1$}}
\put(0.35,-1){\makebox(0,0){#2}} \put(0.35,1){\makebox(0,0){#3}}}

\newcommand{\SEpolt}[5]
{\qbezier(0,-1.5)(.2025,-1.6489)(0.3420,-1.4397)
\qbezier(0.3420,-1.4397)(0.4167,-1.2217)(0.6428,-1.266)
\qbezier(0.6428,-1.266)(0.8937,-1.2499)(0.866,-1)
\qbezier(0.866,-1)(0.7831,-0.7850)(0.9848,-0.6736)
\qbezier(1,-0.5)(1.1,-0.5)(0.9848,-0.6736)
\qbezier(1,0.5)(1.1,0.5)(0.9848,0.6736)
\qbezier(0.9848,0.6736)(0.7831,0.7850)(0.866,1)
\qbezier(0.866,1)(0.8937,1.2499)(0.6428,1.266)
\qbezier(0.6428,1.266)(0.4167,1.2217)(0.3420,1.4397)
\qbezier(0.3420,1.4397)(.2025,1.6489)(0,1.5)
\put(1,0){\VPloopLRt{#1}{#2}}
\put(0.87,-1){\VectorUp}
\put(0.67,1.23){\Vector}
\put(1.3,-1){\makebox(0,0){#3}}
\put(1,0.5){\circle*{0.15}}
\put(1,-0.5){\circle*{0.15}}
\put(0,1.5){\circle*{0.15}}
\put(0,-1.5){\circle*{0.15}}
\put(-0.35,-1.5){\makebox(0,0){#4}}
\put(-0.35,1.5){\makebox(0,0){#5}}}

\newcommand{\SEpoltNA}[5]
{\qbezier(0,-1.5)(.2025,-1.6489)(0.3420,-1.4397)
\qbezier(0.3420,-1.4397)(0.4167,-1.2217)(0.6428,-1.266)
\qbezier(0.6428,-1.266)(0.8937,-1.2499)(0.866,-1)
\qbezier(0.866,-1)(0.7831,-0.7850)(0.9848,-0.6736)
\qbezier(1,-0.5)(1.1,-0.5)(0.9848,-0.6736)
\qbezier(1,0.5)(1.1,0.5)(0.9848,0.6736)
\qbezier(0.9848,0.6736)(0.7831,0.7850)(0.866,1)
\qbezier(0.866,1)(0.8937,1.2499)(0.6428,1.266)
\qbezier(0.6428,1.266)(0.4167,1.2217)(0.3420,1.4397)
\qbezier(0.3420,1.4397)(.2025,1.6489)(0,1.5)
\put(1,0){\circle{1}}
\put(0.87,-1){\VectorUp}
\put(0.67,1.23){\Vector}
\put(1.3,-1){\makebox(0,0){#3}}
\put(1,0.5){\circle*{0.15}}
\put(1,-0.5){\circle*{0.15}}
\put(0,1.5){\circle*{0.15}}
\put(0,-1.5){\circle*{0.15}}
\put(-0.35,-1.5){\makebox(0,0){#4}}
\put(-0.35,1.5){\makebox(0,0){#5}}}

\newcommand{\SEpol}[5]
{\qbezier(0,-1.5)(.2025,-1.6489)(0.3420,-1.4397)
\qbezier(0.3420,-1.4397)(0.4167,-1.2217)(0.6428,-1.266)
\qbezier(0.6428,-1.266)(0.8937,-1.2499)(0.866,-1)
\qbezier(0.866,-1)(0.7831,-0.7850)(0.9848,-0.6736)
\qbezier(1,-0.5)(1.1,-0.5)(0.9848,-0.6736)
\qbezier(1,0.5)(1.1,0.5)(0.9848,0.6736)
\qbezier(0.9848,0.6736)(0.7831,0.7850)(0.866,1)
\qbezier(0.866,1)(0.8937,1.2499)(0.6428,1.266)
\qbezier(0.6428,1.266)(0.4167,1.2217)(0.3420,1.4397)
\qbezier(0.3420,1.4397)(.2025,1.6489)(0,1.5)
\put(1,0){\VPloopLR{#1}{#2}}
\put(0.87,-1){\VectorUp}
\put(0.67,1.23){\Vector}
\put(1.3,-1){\makebox(0,0){#3}}
\put(1,0.5){\circle*{0.15}}
\put(1,-0.5){\circle*{0.15}}
\put(0,1.5){\circle*{0.15}}
\put(0,-1.5){\circle*{0.15}}
\put(-0.35,-1.5){\makebox(0,0){#4}}
\put(-0.35,1.5){\makebox(0,0){#5}}}